\documentclass[prd,showpacs,nofootinbib,preprintnumbers]{revtex4}
\usepackage{amsmath} \usepackage{graphicx} \usepackage{amsfonts}
\usepackage{array} \usepackage{amsthm} \usepackage{bm}

\usepackage{latexsym}
\newcommand{\e}{{\rm e}}

\newcommand{\lb}{\label}

\newcommand{\bw}{\begin{widetext}}
\newcommand{\ew}{\end{widetext}}
\newcommand{\be}{\begin{equation}}
\newcommand{\ee}{\end{equation}}
\newcommand{\bea}{\begin{eqnarray}}
\newcommand{\eea}{\end{eqnarray}}
\newcommand{\nn}{\nonumber}

\begin{document}
\begin{flushright}DTP-MSU/07-25
\end{flushright}

\title{Generating technique for $U(1)^3\;  5D$ supergravity}

\author{Dmitri V. Gal'tsov} \email{galtsov@phys.msu.ru}
\affiliation{Department of Theoretical Physics, Moscow State
University, 119899, Moscow, Russia}

\author{Nikolai G. Scherbluk} \email{shcherbluck@mail.ru}
\affiliation{Department of Theoretical Physics, Moscow State
University, 119899, Moscow, Russia}

\date{\today}

\begin{abstract}
We develop generating technique for solutions of $U(1)^3\; 5D$
supergravity via dimensional reduction to three dimensions. This
theory, which recently attracted attention in connection with black
rings, can be viewed as consistent truncation of the $T^6$
compactification of the eleven-dimensional supergravity. Its further
reduction to three dimensions accompanied by dualisation of the
vector fields leads to $3D$ gravity coupled sigma model on the
homogeneous space $SO(4,4)/SO(4)\times SO(4)$ or
$SO(4,4)/SO(2,2)\times SO(2,2)$ depending on the signature of the
three-space. We construct a $8\times 8$ matrix representation of
these cosets  in terms of lower-dimensional blocks. Using it we
express solution generating transformations in terms of potentials
and identify those preserving asymptotic conditions relevant to
black holes and black rings. As an application we derive the doubly
rotating black hole solution with three independent charges.
Suitable contraction of the above cosets is used to construct a new
representation of the coset $G_{2(2)}/(SL(2,R)\times SL(2,R))$
relevant for minimal five-dimensional supergravity.

\end{abstract}


\maketitle

\section{Introduction}
Since the discovery of black rings \cite{er} in five dimensions, a
variety of solution generation methods were developed  to derive
these solutions in a regular way and to construct  their
generalizations
\cite{elva2,RE,yaza,har1,im,tn,ko,pome1,ak,mi,tmy,ef,posen,ekri,yaza1,
Elvang:2007hs,Elvang:2007hg}. These methods allow to find solutions
possessing a certain number of isometries and they can be combined
into three groups: i) applications of T-duality symmetries acting on
scalar and vector fields in any dimensions, ii) the derivation of
three-dimensional sigma models on homogeneous spaces, iii) further
reduction to two dimensions to apply soliton techniques. Usually the
third approach involves the second one as an intermediate step.
Dimensional reduction of higher-dimensional supergravity theories to
three dimensions accompanied by dualisation of the vector fields
leads to the enhanced U-duality symmetries (hidden symmetries) which
contain transformations useful for generating purposes. So far  only
a restricted class of hidden symmetries of five-dimensional
supergravity  (the vacuum $SL(3,R)$ subgroup \cite{ms,beruf}) was
applied to the black ring problems. Nevertheless, an application of
the level iii) technique based on this subgroup has led to
impressive new results for {\em vacuum} black ring solutions
\cite{har1,im,tn,ko,pome1,ak,mi,tmy,ef,posen,ekri}.

For {\em charged} black rings, only the T-duality at the level i)
was used until recently to generate such solutions from the
uncharged ones. To proceed further to the level ii) one has to
specify the five-dimensional lagrangian containing vector fields.
Pure Einstein-Maxwell theory in five dimensions fails to produce
three-dimensional sigma-model on a symmetric target space. Adding
the Chern-Simons term, as prescribed by the minimal five-dimensional
supergravity \cite{cre,chani}, one obtains a more symmetric
three-dimensional sigma-model with an exceptional group $G_{2(2)}$
acting as the target space isometry
\cite{CFG,BoCa,dWVvP,dWvP,mizo,cjlp,mizo2,pos1,pos,gnpp,Nikolai,dhw,BePi}.
The corresponding generating technique was recently developed in
\cite{bccgsw,gc07}. It amounts to using the $7\times 7$ matrix
representation of the coset $G_{2(2)}/(SL(2,R)\times SL(2,R))$ and
opens a way to construct the most general five-parametric black ring
solution of the minimal five-dimensional supergravity as well as its
possible generalizations such as charged Saturns.

The purpose of this paper is to generalize the same approach to
$U(1)^3$  five-dimensional supergravity with three vector fields.
This theory  can be regarded as a truncated toroidal
compactification of the  $11D$  supergravity: \be \label{ans11}
I_{11} = \frac{1}{16\pi G_{11}}\int\left(R_{11}\star_{11}
\mathbf{1}-\frac12 F_{[4]}\wedge \star_{11} F_{[4]} -
\frac16F_{[4]}\wedge F_{[4]} \wedge A_{[3]}\right), \ee where $
F_{[4]} = dA_{[3]} $, according to an ansatz \bea\label{met11}
ds_{11}^2 &=& ds^2_{5} + X^1 \left( dz_1^2 + dz_2^2 \right) + X^2
\left( dz_3^2 + dz_4^2 \right) + X^3 \left( dz_5^2 + dz_6^2
\right),
\\
A_{[3]} &=& A^1 \wedge dz_1 \wedge dz_2 + A^2 \wedge dz_3 \wedge
dz_4 + A^3 \wedge dz_5 \wedge dz_6 \,. \nn \eea Here $z^i,\
i=1,\ldots,6$ are the coordinates parameterizing the torus $T^6$.
The three scalar moduli $X^I,(I=1,2,3)$ and the three one-forms
$A^I$ depend only on the five coordinates entering $ds^2_5$. The
moduli $X^I$ satisfy the constraint $X^1X^2X^3=1$,  implying that
the five-dimensional metric $ds_5^2$ is the Einstein-frame metric.
The reduced five-dimensional action reads: \bea\label{L5} I_5 &=&
\frac{1}{16 \pi G_5} \int \left( R_5 \star_5 \mathbf{1} - \frac12
G_{IJ} dX^I \wedge \star_5 dX^J - \frac12G_{IJ} F^I \wedge \star_5
F^J -
\frac{1}{6} \delta_{IJK} F^I \wedge F^J \wedge A^K \right),\nn \\
G_{IJ}&=&{\rm diag}\left((X^1)^{-2},\ (X^2)^{-2},\
(X^3)^{-2}\right),\quad F_{I}=dA_{I},\quad I,J,K=1,2,3,\nn
 \eea
where the Chern-Simons coefficients $\delta_{IJK}=1$ for the indices
$ I,J,K $ being a permutation of 1, 2, 3, and zero otherwise.
Supersymmetric solutions to this theory were studied in a number of
papers
\cite{MT,3charge,3chargepert,EMT,EE,BK,bena,EEMR1,EEMR2,BW,JGJG2}.
The most general ring solution to this theory constructed so far
\cite{EEF} is a family of non-supersymmetric rings parameterized by
three conserved charges $Q_I$, three dipole charges $q_I$, and a
radius of $S^1$, with the mass $M$ and the two angular momenta $
J_\psi,J_\phi$ being some functions of these seven free parameters.
An existence of a larger family of non-supersymmetric black rings
with nine independent parameters $(M,J_\psi,J_\phi,Q_{I},q_{I})$ is
expected, such that reduces to the solutions of
\cite{EEMR1,EEMR2,BW,JGJG2} in the supersymmetric limit. The
generating technique developed in the present paper provides a
sufficient number of parameters to construct the nine-parametric
solution.

It is worth noting that the ansatz (\ref{met11}) is far from being
the general toroidal compactification of the  $11D$  supergravity.
The generic toroidal reduction leads to the five-dimensional theory
with 27 vector fields and 42 scalar moduli, parameterizing a coset
$E_{6(6)}/USp(8)$. Correspondingly, the general black ring must
contain 27 conserved charges and 27 dipole charges. More accurate
analysis \cite{larsen} shows that 24 conserved charges can be
generated from the above three by duality transformations, while the
number of independent dipole charges is  15 (the number of
independent four-cycles of $T^6$).

Contraction of the above theory to minimal $5D$ supergravity is
effected via an identification of the vector fields:
$$A^1=A^2=A^3=\frac{1}{\sqrt{3}}A,$$ and freezing out  the moduli:
$X^1=X^2=X^3=1$. This leads to the Lagrangian   \be \mathcal{L}_5
= R_5 \star_5 \mathbf{1} - \frac12 F\wedge \star_5 F -
\frac{1}{3\sqrt{3}} F\wedge F \wedge A.\nn \ee In this case our
results go back to those of the Refs. \cite{bccgsw,gc07}. However,
our matrix representation of the coset $SO(4,4)/SO(4)\times SO(4)$
leads upon this contraction to a new representation of the coset
$G_{2(2)}/(SL(2,R)\times SL(2,R))$,  different and somewhat
simpler than given in \cite{bccgsw,gc07}.

\section{3D sigma-model}
\subsection{Dimensional reduction}
To perform dimensional reduction of the $11D$ theory to three
dimensions we follow an approach of Ref. \cite{cjlp1} (keeping all
basic notations of that paper) which has an advantage to provide the
roots of the hidden symmetry group directly in terms of the so
called {\em dilaton vectors} (coefficients in the dilaton
exponentials entering the reduced action). For this purpose we go
back to eleven dimensions and consider compactification of
eleven-dimensional supergravity on the torus $T^8=T^6\times T^2$
parameterized by coordinates $z^1,\ldots, z^8$. It will be
convenient to distinguish the  six coordinates on the torus $T^6$,
corresponding to the reduction  to five dimensions,
$z^i,\,i=1,\ldots,6$, from those on $T^2$, corresponding to the
reduction from five to three dimensions, which will be denoted by
elder indices $p,q=7,8$. The decomposition of the eleven-dimensional
metric in terms of the five-dimensional and three-dimensional
metrics incorporating a diagonal ansatz (\ref{met11}) on $T^6$
(sector $z^i$) and the KK ansatz  on $T^2$ (sector $z^p$)  then
reads in the notation of \cite{cjlp1}: \vspace{-.2cm}
\bea
 ds_{11}^2
 &=&  \e^{\vec{s}\cdot\vec{\sigma}} ds_{5}^2+
 \sum_{k=1}^6 \e^{2\vec{\gamma}_k\cdot\vec{\sigma}}
(dz^k)^2 \\ \vspace{-.3cm} &=& \e^{\vec{s}\cdot\vec{\theta}}
ds_{3}^2 + \e^{\vec{s}\cdot \vec{\sigma}}\left[
\e^{2\vec\gamma_7\cdot \vec\varphi}(dz^7+\mathcal{A}^7+\chi
dz_8)^2+\kappa
 \e^{2\vec\gamma_8\cdot\vec\varphi}(dz^8+\mathcal{A}^8)^2
 \right]+\sum_{k=1}^6 \e^{2\vec{\gamma}_k\cdot\vec{\sigma}}
 (dz^k)^2\nn,
 \eea
where  $\mathcal{A}^7,\mathcal{A}^8$ are three-dimensional
Kaluza-Klein one-forms from the reduction of the five-dimensional
metric on $T^2,\; \chi$ is an axion arising in the reduction of the
four-dimensional one-form $\mathcal{\hat A}^7$ on the second cycle
of $T^2$: $\mathcal{\hat A}^7=\mathcal{A}^7 +\chi dz^8$ (in the
notation of the Ref. \cite{cjlp1} $\chi=\mathcal{A}^7_8$), the
factor $\kappa=\pm 1$ is responsible for the signature: $
 \kappa=1$ for a space-like $z^8$, and $\kappa=-1$ for a time-like $z^8$ .
The eight-dimensional dilaton $\vec{\theta}$ is split into the sum
of the $T^6$ and $T^2$ components: \be
\vec\theta=\vec\sigma+\vec\varphi,\quad
 \vec\sigma=(\sigma_1,...,\sigma_6,0,0),\quad
 \vec\varphi=(0,...,0,\varphi_1,\varphi_2),\nn
\ee and the  dilaton vectors can be presented as \bea \lb{sf}
 \vec s&=& (s_1,\ldots,s_8),\quad s_k = \sqrt{2/((10-k)(9-k))},\quad
 \vec{\gamma}_k=\frac12(\vec s-\vec{f}_k),\nn
 \\
\vec f_k &=& \Big(\underbrace{0,0,\ldots, 0}_{k-1}, (10-k) s_k,
s_{k+1}, s_{k+2}, \ldots, s_8\Big), \quad k=1...8. \lb{sf}
 \eea
The relation between the dilatons and the moduli $X^I$ with account
for the constraint $X^1X^2X^3=1$ is given by:
\bea \label{relation_on_dil_vec1}
 \vec{s}\cdot\vec{\varphi}&=&\frac{1}{\sqrt3}\varphi_1+\varphi_2,\nn
 \\   \vec{s}\cdot\vec{\sigma}&=&0\ \Leftrightarrow \ X^1X^2X^3=1,
 \eea \vspace{-.3cm}
 \bea\label{relation_on_dil_vec2}
X^1&=&\e^{2\vec{\gamma}_1\cdot\vec\sigma}=\e^{2\vec{\gamma}_2\cdot\vec\sigma},\nn
\\
X^2&=&\e^{2\vec{\gamma}_3\cdot\vec\sigma}=\e^{2\vec{\gamma}_4\cdot\vec\sigma},
\\
X^3&=&\e^{2\vec{\gamma}_5\cdot\vec\sigma}=\e^{2\vec{\gamma}_6\cdot\vec\sigma},\nn
\eea
 \vspace{-.8cm}
 \be
2\vec\gamma_7\cdot\vec\varphi=-\frac{2}{\sqrt3}\varphi_1,\quad
2\vec\gamma_8\cdot\vec\varphi=\frac{1}{\sqrt3}\varphi_1-\varphi_2,\nn\\
 \ee
 \vspace{-.6cm}
 \be\label{qu_form_alpha}
 (\partial\vec{\sigma})^2
=\sum_{I=1}^{3}\left(\frac{\partial X^I}{X^I}\right)^2.
  \ee

To rewrite the ansatz for the eleven-dimensional three-form
potential $A_{[3]}$ in the notation of the Ref. \cite{cjlp1} we will
use the pairwise indices $ii', jj',\ldots$  taking three values
$ii'=(12,34,56)$, together with the indices on $T^2$: $p=7,8,$ so
that \vspace{-.2cm}\be A_{[3]}= \frac12A_{ii'}\wedge dz^i\wedge
dz^{i'} + \frac16 A_{ii'p}dz^i\wedge dz^{i'} \wedge dz^p.\nn
 \ee
Here the one-forms $ A_{12}=A^1, A_{34}=A^2, A_{56}=A^3$   are the
pull-back of the five-dimensional one-forms introduced in
(\ref{met11}) onto the three-space (assuming $A_{ii'}=-A_{i'i}$),
and the scalars $A_{ii'p}=(A_{127}, A_{347}, A_{567}, A_{128},
A_{348}, A_{568})$ are  axions arising in the reduction of the
five-dimensional one-forms on $T^2$.

Using the result of the Ref. \cite{cjlp1}, we obtain the following
three-dimensional Lagrangian (for $\kappa=1$):  \be \label{L3}
 e_3^{-1}{\cal L}_3 =   R_3 -\frac12  (\partial\vec \theta)^2
 -\frac14 \sum_{i<i'} \e^{\vec
a_{ii'}\cdot \vec{\theta}}  (F_{ii'})^2 -\frac14  \, \sum_p
\e^{\vec b_p\cdot \vec\varphi}\, ({\cal F}^p)^2 -\frac12
\!\!\sum_{i<i',p} \e^{\vec a_{ii'p} \cdot\vec \theta}\,
(F_{ii'p})^2 - \frac12\e^{\vec b_{78}\cdot \vec\varphi}\,
(\partial\chi)^2 +e_3^{-1}{\cal L}_{CS},\ee where the field
strength two-forms are defined as
 \bea {\cal
F}^7&=&d{\cal A}^7 -d\chi\wedge {\cal A}^8,\quad {\cal F}^8=d{\cal
A}^8,\nn\\
F_{ii'}&=&dA_{ii'}-dA_{ii'p}\gamma^p{}_q\wedge {\cal A}^q,\quad
 \gamma^7{}_7=\gamma^8{}_8=1,\quad \gamma^7{}_8=-\chi,\nn\\
 F_{ii'p}&=&\gamma^q{}_pdA_{ii'q}.\nn
 \eea
The newly introduced dilaton vectors are related to the quantities
defined in (\ref{sf}) via \bea\label{dilaton_vectors}
 \vec a_{ii'} &=& \vec f_i + \vec f_{i'} - 3\vec s,\quad\quad
 \vec a_{ii'p} = \vec f_i + \vec f_{i'} + \vec f_{p} - 3\vec s,
  \\
 \vec b_p&=&-\vec f_p,\qquad\qquad\qquad\  \vec b_{78} = \vec f_8-\vec
 f_7.\nn
 \eea
Finally, the Chern-Simons term reads: \be {\cal
L}_{CS}=-\frac{1}{144} \epsilon^{ii'pkk'qjj'}dA_{ii'p}\wedge
dA_{kk'q}\wedge {A}_{jj'},\nn \ee where an eight-dimensional ($T^8$)
Levi-Civita symbol is used. \par In our truncated toroidal reduction
the six-dimensional part of the dilaton $\vec\sigma$  is effectively
two-dimensional in view of the four constraints
 (\ref{relation_on_dil_vec1})-(\ref{relation_on_dil_vec2}). So, together
 with the two-dimensional $T^2$-dilaton $\vec\varphi$, one gets
instead of the 8-dimensional vector $\vec\theta$ the
four-dimensional vector $\vec\phi=(\phi_1,
 \phi_2,\phi_3,\phi_4)$ with the following components:
\bea
 \phi_1&=&\frac{1}{\sqrt2}\left(-\ln(X^3)+\frac{1}{\sqrt3}
 \varphi_1+\varphi_2\right),\quad
   \phi_2=\frac{1}{\sqrt2}\left(\ln(X^3)-\frac{1}{\sqrt3}
   \varphi_1+\varphi_2\right),\label{phi1&phi2}
 \\
 \phi_3&=&\frac{1}{\sqrt2}\left(\ln(X^3)+\frac{2}{\sqrt3}
 \varphi_1\right),\qquad\qquad
 \phi_4=\frac{1}{\sqrt2}\,\ln\frac{X^1}{X^2}.\label{phi3&phi4}
 \eea
 It has the same norm: \vspace{-.5cm}
 \be
  \sum_{k=1}^8\partial \theta_k^2= \sum_{k=1}^4\partial \phi_k^2,\nn
 \ee
and the corresponding exponents convert to the four-dimensional ones
as follows: \be
    \e^{\vec{a}_{12}\cdot\vec{\theta}}\rightarrow
    \e^{\vec{a}_{12}\cdot\vec{\phi}},\quad
    \e^{\vec{b}_{7}\cdot\vec{\varphi}}\rightarrow
    \e^{\vec{b}_{7}\cdot\vec{\phi}},\ldots,
    \quad
    \vec\phi=(\phi_1,\phi_2,\phi_3,\phi_4),\nn
 \ee
while the four-dimensional dilaton coefficient vectors in the new
basis read: \bea
 \vec{a}_{12}&=&\sqrt2(-1,0,0,-1),\qquad\qquad\
 \vec{a}_{34}=\sqrt2(-1,0,0,1),\qquad\qquad\quad
 \vec{a}_{56}=\sqrt2(0,-1,-1,0),\nn
 \\
 \vec{a}_{127}&=&\sqrt2(0,0,1,-1),\qquad\qquad\ \ \
 \vec{a}_{347}=\sqrt2(0,0,1,1),\qquad\qquad\quad
 \vec{a}_{567}=\sqrt2(1,-1,0,0),\nn
 \\
 \vec{a}_{128}&=&\sqrt2(0,1,0,-1),\qquad\qquad\ \
 \
 \vec{a}_{348}=\sqrt2(0,1,0,1),\qquad\qquad\quad
 \vec{a}_{568}=\sqrt2(1,0,-1,0),\nn
 \\
 \vec{b}_{7}&=&\sqrt2(-1,0,-1,0),\qquad\qquad\quad
 \vec{b}_{8}=\sqrt2(-1,-1,0,0),\qquad\qquad\
 \vec{b}_{78}=\sqrt2(0,1,-1,0).\nn
 \eea
\subsection{$T^2$-covariant form}
It will be useful to perform reduction  on a torus $T^2$ in the
$2D$-covariant form decomposing the five-metric as
  \be
 ds_5^2=\lambda_{pq}(dz^p+a^p)(dz^q+a^q)-\kappa\tau^{-1}ds_3^2,\nn
 \ee
where  the $2D$ metric $\lambda_{pq}\, (p=7,8)$ is introduced:   \be
 \lambda=\e^{-\frac{2}{\sqrt3}\varphi_1}\left( \begin{array}{cc}
1&
\chi \\
\chi & \chi^2+\kappa \e^{\sqrt{3}\varphi_1-\varphi_2}\end{array}
\right),\qquad  \det\lambda\equiv -\tau=\kappa
 \e^{-\frac{1}{\sqrt3}\varphi_1-\varphi_2},\label{def_lambda}
 \ee
 and $a^p$ are the Kaluza-Klein one-forms:
$
 a^7={\cal A}^7-\chi{\cal A}^8,\quad a^8={\cal
 A}^8.\nn
 $
For the moduli $X^{I}$ we have the following expressions in terms of
$\tau$ and the dilatons $\phi_1, \phi_4$:
\be (X^1)^2=\e^{\sqrt2\phi_4}(X^3)^{-1},\quad
 (X^2)^2=\e^{-\sqrt2\phi_4}(X^3)^{-1},\quad
 X^3=\tau^{-1}\e^{-\sqrt2\phi_1}\label{X_via_phi}.
 \ee
Using the relations (\ref{phi1&phi2})-(\ref{phi3&phi4}) we can
rewrite the metric $G_{IJ}$ of the moduli space as
  \be
   G_{IJ}=-\frac{\kappa}{\tau}{\rm diag}(\e^{\vec a_{12}\cdot\vec \phi},
   \e^{\vec a_{34}\cdot\vec \phi},\e^{\vec a_{56}\cdot\vec
   \phi}).\label{G_IJ}
   \ee

\subsection{Dualisation }
To obtain a purely scalar $3D$ Lagrangian we have to perform {\em
dualisation} of the 2-forms $F_{ii'}$ and $\mathcal{F}^p$. First of
all we change notation from that of the Ref. \cite{cjlp1} replacing
the pairwise indices $ii'=12, 34, 56$ by a capital Roman index $I=1,
2, 3$, and relabeling the axions similarly: \be F^I=(
F_{12},F_{34},F_{56}),\quad
 u^I=(A_{127},A_{347},A_{567}),\quad v^I=
(A_{128},A_{348},A_{568}).\nn\ee It is important to realize that the
indices $I$ are the vector indices in the moduli space endowed with
the metric $G_{IJ}$. We also combine $u^I$ and $v^I$ into the
$T^2$-covariant doublet $\psi^I_p=(u^I,\,v^I)$  with the index $p$
relative to the metric $\lambda_{pq}$, or, in the matrix form, $
 \psi^I=\left( \begin{array}{cc}  u^I\\
 v^I \end{array} \right).
$
In what
 follows the summation over all the repeated indices is understood.
 In this notation the field strength tensors will read:
 \be
 F^I=dA^I-d\psi_p^I\wedge a^p,\quad {\cal F}^7=da^7+\chi
 da^8,\quad {\cal F}^8=da^8.\label{def_F_via_a_psi}
 \ee

To perform dualisation along the lines of  \cite{cjlp1} we introduce
into the lagrangian  (\ref{L3}) three Lagrange multipliers $\mu_I$
ensuring the Bianchi identities for the two-forms
$F^I-\psi_p^Ida^p=dA^I-d(\psi_p^Ia^p)$ and two Lagrange multipliers
$\omega_p$ ensuring the Bianchi identities for the two-forms
$da^p=\gamma^p{}_q\mathcal{F}^q$. We also rewrite the Chern-Simons
term as follows (see Eq. (3.29) in \cite{cjlp1}):
$$
 {\cal
L}_{CS}=\frac12\delta_{IJK}\epsilon^{pq}(d\psi_p^I\psi_q^J\wedge
F^K+\frac13d\psi_p^I\psi_q^J\psi_r^K\gamma^r{}_s\wedge{\cal F}^s),
$$
with $\epsilon^{pq}=-\epsilon^{qp},\,\epsilon^{78}=1$. Integrating
by parts we can present the lagrangian (\ref{L3}) as
 \bea
{\cal L}_3&=&R_3\star\mathbf{1}-\frac12 \star d\vec\phi\wedge
d\vec\phi-\frac12 \!\!\sum_{i<i',p} \e^{\vec a_{ii'p} \cdot\vec
\phi}\, \star F_{ii'p}\wedge F_{ii'p}-\frac{\tau\kappa}{2}
G_{IJ}\star F^I\wedge
 F^J+G_I\wedge F^I\nn\\
  &-&\frac12 \e^{\vec b_7\cdot\vec\phi}\star{\cal F}^7\wedge {\cal
 F}^7
 -\kappa\frac12 \e^{\vec b_8\cdot\vec\phi}\star{\cal F}^8\wedge {\cal
F}^8+G_p\wedge {\cal F}^p\nn,
 \eea
where the one-forms $G_I, G_p$ are related to the scalars
$\mu_I,\,\omega_p$ as follows:
 \bea
  &&G_I=d\mu_I+\frac12\delta_{IJK}d\psi_p^J
 \psi_q^K\epsilon^{pq},\nn\\
 && G_7=V_7,\quad G_8=V_8-\chi V_7,\nn\\
 && V_p=d \omega_p- \psi_p^I  \Big( d \mu_I+\frac16 \delta_{IJK}d\psi_q^J
 \psi_r^K\epsilon^{qr}\Big),\nn
  \eea
Then, eliminating the initial two-forms $F^I,\,{\cal F}^p$ via the
equations of motion
 \be
 F^I=\tau^{-1}G^{IJ}\star G_J,\quad
 {\cal F}^7=-\kappa \e^{-\vec b_7\cdot\vec\phi}\star G_7,\quad {\cal F}^8=
 -\e^{-\vec b_8\cdot\vec\phi}\star
 G_8,\label{dual_eq}
 \ee
 we obtain the lagrangian in the dual terms:
 \bea
{\cal L}_3&=& R_3\star\mathbf{1}-\frac12 \star d\vec\phi\wedge
d\vec\phi-\frac12 \!\!\sum_{i<i',p} \e^{\vec a_{ii'p} \cdot\vec
\phi}\, \star F_{ii'p}\wedge F_{ii'p}\nn\\ &+&\frac12\tau^{-1}
G^{IJ}\star G_I\wedge G_J -\kappa\frac12 \e^{-\vec
b_7\cdot\vec\phi}\star G_7\wedge G_7-\frac12\e^{-\vec
b_8\cdot\vec\phi} \star G_8\wedge
 G_8,\nn
 \eea
where $G^{IJ}$ is the inverse moduli metric $G_{IJ}$. Note that
the signs in the dilaton exponents were inverted under
dualisation. The Eqs. (\ref{def_F_via_a_psi}) together with the
relations
$
  \e^{\vec b_7\cdot\vec\phi}=-\kappa\tau \lambda_{77},\ \e^{\vec b_8\cdot\vec\phi}=
 \tau(\chi\lambda_{78}-\lambda_{88}),
 $
which follow from the definitions
(\ref{phi1&phi2})-(\ref{phi3&phi4}) and (\ref{def_lambda}), enable
us to rewrite the Eqs.(\ref{dual_eq}) as the dualisation equations
covariant with respect to all indices:
\vspace{-.1cm} \bea
 && \tau\lambda_{pq}da^q=\star V_p,\nn\\
  dA^I&=&d\psi_p^I\wedge a^p+\tau^{-1}G^{IJ}\star G_J,\nn
  \eea
or, explicitly: \vspace{-.4cm}
\bea \lb{dua1}
 &&\lambda_{pq}\partial^{{[}i} a^{j{]q}}=\frac{1}{2\tau \sqrt
 h}\epsilon^{ijk}\Bigg[\partial_k \omega_p- \psi_p^I  \left(\partial_k \mu_I+
 \frac16 \delta_{IJK}\partial_k\psi_r^J
 \psi_s^K\epsilon^{rs}\right)\Bigg], \\
 &&\partial^{{[}i} A^{j{]}I}=a^{p{[}j}\partial^{i{]}}
 \psi_p^I+\frac{1}{2\tau\sqrt{h}}\epsilon^{ijk}G^{IJ}\left(\partial_k\mu_J+
 \frac12\delta_{JKL}\partial_k\psi_p^K
 \psi_q^L\epsilon^{pq}\right), \lb{dua2}
 \eea
where  the antisymmetrization  is assumed with $1/2$.

Combining all the above formulas we can present the dualised action
as that of a $3D$ gravity coupled sigma model: \be
 I_3=\frac{1}{16\pi G_3}\int \sqrt{|h|}\left(R_3-{\cal G}_{AB}
 \frac{\partial\Phi^A}{\partial
x^i}\frac{\partial\Phi^B}{\partial x^j}h^{ij}\right)d^3x,\nn
 \ee
where $h^{ij}$ is the inverse metric of the three-space, $R_3$ is
the corresponding Ricci scalar and ${\cal G}_{AB}(\Phi^A)$ is the
 metric of the target space parameterized by sixteens scalar variables
 $
 \Phi^A=(\vec\phi,  \psi^I, \mu_I,\chi,\omega_p),\nn
 $
which can be read off from the following line element:
\begin{eqnarray}\label{TS_SO44}
dl^2&=&{\cal{G}}_{AB}d\Phi^Ad\Phi^B
\\ \nonumber
&=& \frac12 \Big((d\vec\phi)^{2} +\kappa \e^{\sqrt {2}
(\phi_1+\phi_4)} (G_1)^2+\kappa \e^{\sqrt {2}( \phi_1-\phi_4)}
(G_2)^2+ \kappa \e^{\sqrt{2}(\phi_3+\phi_2)} (G_3)^2
\\ \nonumber
&+& \e^{ \sqrt{2}(\phi_3-\phi_4)} (du^1)^{2}+\e^{\sqrt
{2}(\phi_4+\phi_3)} (du^2)^{2}+\e^{\sqrt {2}(\phi_1-\phi_2)}
(du^3)^{2}
\\ \nonumber
&+&\kappa \e^{\sqrt {2} (\phi_2-\phi_4)} \left(dv^1 -\chi
du^1\right) ^{2}+\kappa \e^{\sqrt{2}( \phi_4+\phi_2)} \left( dv^2
-\chi du^2\right) ^{2}+\kappa \e^{\sqrt {2} (\phi_1-\phi_3)}
\left( dv^3 -\chi du^3 \right) ^{2}
\\ \nonumber
&+&\kappa \e^{\sqrt {2} (\phi_1+\phi_3)} (G_7)^{2}+
 \e^{\sqrt{2}(\phi_1+\phi_2)}(G_8)^2+ \kappa \e^{\sqrt
{2}(\phi_2-\phi_3)}d\chi^{2} \Big).
\end{eqnarray}
This line element can be more concisely rewritten in the
$T^2$-covariant form:
\be
 dl^2\! =\! \frac12
G_{IJ}(dX^IdX^J\!+\!d{{\psi^I}^T}\lambda^{-1}
d\psi^J)\!-\!\frac12\tau^{-1}G^{IJ}G_IG_J\!+\!\frac14 \mathrm{Tr}
\left( \lambda^{-1} d\lambda \lambda^{-1} d\lambda \right)\! +\!
\frac14 \tau^{-2} d\tau^2\! - \!\frac12\tau^{-1} V^T \lambda^{-1}
V.\nn
 \ee

\subsection{Hidden symmetry}
The set of the dilaton vectors $\vec
{a}_{ii'},\vec{a}_{ii'p},\vec{b}_p,\vec{b}_{78}$ is directly
related to  the root system of the isometry algebra of the target
space \cite{cjlp1}. Enumerating them as
 \bea
 -\vec{a}_{ii'}=(\vec{e}_1,\vec{e}_2,\vec{e}_3);\quad
\vec{a}_{ii'7}=(\vec{e}_4,\vec{e}_5,\vec{e}_6);\quad
\vec{a}_{ii'8}= (\vec{e}_7,\vec{e}_8,\vec{e}_9);\quad
-\vec{b}_p=(\vec{e}_{10},\vec{e}_{11});\quad
\vec{b}_{78}=\vec{e}_{12},\nn
 \eea
one can  easily see that these twelve four-dimensional vectors form
the system of positive roots of the algebra $so(8)$. Indeed, from
the relations (\ref{dilaton_vectors}) and the property \vspace{-2mm}
 \be
 \sum_{k=1}^8 \vec{f}_k=9\vec s,\nn
 \ee
 one
can express $\vec{e}_1,\vec{e}_2,\vec{e}_3,
\vec{e}_7,\vec{e}_8,\vec{e}_9,\vec{e}_{10},\vec{e}_{11}$ in terms of
$ \vec{e}_4,\vec{e}_5,\vec{e}_6, \vec{e}_{12}$  as follows: \bea
 \vec e_I&=&\sum_{K\neq I}\vec e_{K+3}+ \vec
e_{12},\quad\
 \vec e_{I+6}=\vec e_{I+3} + \vec e_{12},\nn \\
 \vec e_{10}&=&\sum_{K}\vec e_{K+3}+\vec e_{12},\quad
 \vec e_{11}=\sum_{K}\vec e_{K+3}+2\vec e_{12}.\nn
 \eea
It is clear then that the vectors $\vec e_4,\vec e_5,\vec e_6,\vec
e_{12}$ are the simple roots forming the Dynkin diagram  of
$so(8)$ \cite{Gilmore}.

The signature of the target space is $+16$ for $\kappa=1$
(dimensional reduction in all space-like directions) and $(+8,-8)$
$\kappa=-1$ (one of the reduced dimensions is time-like). Then it
is easy to recognize that the isometry group is  the non-compact
form $SO(4,4)$ of the $SO(8)$, whose Killing metric has the
signature $(-12,+16)$, while the target space is the coset
$SO(4,4)/ SO(4)\times SO(4)$ for $\kappa=1$ and
$SO(4,4)/SO(2,2)\times SO(2,2)$ for $\kappa=-1$.  For these both
symmetric spaces    the scalar curvature  is negative: $${\cal
R}=-96.$$
\par Denoting the four-dimensional Cartan subalgebra of $so(4,4)$ as
$\vec H$, and the generators corresponding to the non-zero roots
$\pm\vec e_k,\, k=1,\ldots, 12$ as $P^{\pm I},\, W_{\pm I},\,
Z_{\pm I},\,\Omega^{\pm p},\,X^{\pm}$, we will have the relations:
\be P^{\pm I}\leftrightarrow {\pm\vec e_I}, \quad W_{\pm
I}\leftrightarrow {\pm\vec e_{I+3}}, \quad Z_{\pm
I}\leftrightarrow {\pm\vec e_{I+6}}, \quad \Omega^{\pm
p}\leftrightarrow {\pm\vec e_{p+3}},\quad X^{\pm} \leftrightarrow
{\pm\vec e_{12}}.\nn\ee The commutators of these generators with
the Cartan subalgebra $\vec H$ read: \bea
  {[}\vec{H},X^{\pm}{]}&=& \pm\vec{e}_{12}X^{\pm},\label{com_H}\nn \\
  {[}\vec{H},\Sigma_{\pm I}^p{]}&=& \pm\vec{e}_{I+3(p-6)}\Sigma_{\pm I}^p,\\
  {[}\vec{H},\Omega^{\pm p}{]}&=& \pm\vec{e}_{p+3}\Omega^{\pm
  p},\nn
 \\
  {[}\vec{H},P^{\pm I}{]}&=& \pm\vec{e}_{I}P^{\pm I}, \nn\eea
where we have arranged  $W_I,\,Z_I$ into a column
vector $ \Sigma_I =\left( \begin{array}{cc} W_I\\
Z_I \end{array} \right).$ The remaining non-zero commutators are
obtained from the relations between the root vectors \bea
 \vec{e}_{I+3}+\vec{e}_{J+6}=\vec{e}_{K},\quad \vec{e}_{I+3}+\vec{e}_{12}
 =\vec{e}_{I+6},
 \quad \vec{e}_{I+3(a+1)}+\vec{e}_{I}=\vec{e}_{a+10}\ (a=0,1),\quad
 \vec{e}_{12}+\vec{e}_{10}=\vec{e}_{11},\nn
 \eea
 where in the first equations $I,J,K$ are all different.
 One finds: 
 \bea
  \label{comm_of_so(4,4)_alg}
  {[}\Sigma_{\pm I}^p,\Sigma_{\pm J}^q{]}&=&\mp \epsilon^{pq}\delta_{IJK}
  P^{\pm
 K},\quad {[}\Sigma_{\mp I}^p,\Sigma_{\pm J}^q{]}=\pm
 \epsilon^{pq}\delta_{IJ}X^{\pm},\nn\\
 {[}X^{\pm},W_{\pm I} {]}&=&\mp Z_{\pm I},\quad {[}X^{\mp},Z_{\pm I} {]}=
 \mp W_{\pm I},\\
 {[}\Sigma_{\pm I}^p,P^{\pm J} {]}&=&\mp \delta^J_I \Omega^{\pm p},\quad
 {[}\Sigma_{\pm I}^p,P^{\mp J} {]}=\pm\epsilon_{.q}^{p} \delta_{IK}
 \delta^{KJL}\Sigma_{\mp L}^q ,\nn \\
  {[}X^{\pm},\Omega^{\pm 7}{]}&=&\mp\Omega^{\pm 8}. \nn
  \eea
We will give the generators of $SO(4,4)$ as differential operators
acting on the target space manifold in what follows.

\section{Coset representative}
\subsection{The strategy}
As a convenient representative of the coset one can choose the upper
triangular matrix ${\cal V}$ which transforms under the global
action of the  symmetry group $ G$ by the right multiplication and
under the local action of the isotropy group $H$ by the left
multiplication: \be {\cal V}\to {\cal V}'=h(\Phi){\cal V}g,\quad
 g\in G, \quad h\in H.\nn
 \ee
Given this representative, one can construct the $H-$invariant
matrix \be
 {\cal M}={\cal V}^T K {\cal V},\label{def_M}
 \ee
where $K$ is an involution matrix invariant under $H$: \be
 h(\Phi)^T K h(\Phi)=K,\label{h_K_h}
 \ee
(dependent on the coset signature parameter $\kappa$).
 Then the transformation of ${\cal M}$ under $G$ will be
\be {\cal M}\to {\cal M}'=g^T {\cal M}g.\label{act_G_on _M}
 \ee
The target space metric (\ref{TS_SO44}) in terms of the matrix
${\cal M}$ will read  \be
 dl^2=-\frac18\texttt{tr}(d{\cal M}d{\cal M}^{-1}).\label{TS_by_M}
 \ee

The desired upper-triangular matrix ${\cal V}$ can be constructed by
an exponentiation of the Borel subalgebra of the Lie algebra of $G$
consisting of the Cartan $H$ and the positive-root $E_{+}$
generators (in what follows we omit the sign $+$ in the indices):
\be
 {\cal V}={\cal V}_H{\cal V}_{E_{+}}={\cal V}_H{\cal V}_{X}
 {\cal V}_{\Psi}{\cal V}_{\Omega}{\cal
 V}_{P},\nn
 \ee
where the matrices ${\cal V}_{H},\ {\cal V}_{X},\ {\cal V}_{\Psi},\
{\cal V}_{\Omega},\ {\cal V}_{P}$ are the exponentials:
\bea\label{def_Nu}
 {\cal V}_H&=&\e^{\frac12 \vec{\phi}\cdot \vec H},\nn\\
 {\cal V}_X&=&\e^{\chi X},\nn\\
 {\cal V}_{\Psi}&=&\e^{\psi^I \Sigma_I},\\
 {\cal V}_{\Omega}&=&\e^{\omega_p \Omega^p},\nn\\
 {\cal V}_P&=&\e^{\mu_I P^I}.\nn
 \eea
Using (\ref{TS_by_M}), one can rewrite the   target space metric in
terms of the  matrix current  ${\cal J}=d{\cal V}{\cal V}^{-1}$ as
follows: \be
 dl^2=\frac14\texttt{tr}(\mathcal{J}^2)+\frac14\texttt{tr}
 (\mathcal{J}^{\texttt{T}}K\mathcal{J}K^{-1}).\nn
 \ee
Using the Eqs.(\ref{def_Nu}) and the commutators (\ref{com_H}) and
(\ref{comm_of_so(4,4)_alg}) for the positive-root generators, one
can show that the matrix current one-form ${\cal J}$ is spanned by
the Borel subalgebra generators as follows: \bea
 \mathcal{J}= d{\cal V}\, {\cal V}^{-1}&=& \frac12
d\vec\phi\cdot\vec H + \e^{\frac12 \vec
{e}_{12}\cdot\vec\phi}\,d\chi\,X +\sum_{I} \e^{\frac12 \vec
{e}_{I+3}\cdot\vec\phi}\, du^I\,W_I+ \sum_{I} \e^{\frac12
\vec {e}_{I+6}\cdot\vec\phi}\,  \left(dv^I -\chi du^I\right)\, Z_{I}\nn \\
&& + \sum_{p}\e^{\frac12\vec {e}_{p+3}\cdot \vec\phi}\, G_{p}\,
\Omega^p +\sum_{I} \e^{\frac12\vec {e}_{I}\cdot \vec\phi}\,
G_I\,P^I.\nn
 \eea
\subsection{Matrix representation}
We use the $8\times 8$ matrix representation of the $so(4,4)$
algebra given in the Appendix A. The exponentiation of the Borel
subalgebra gives the coset representatives in the following block
form:
\be
 {\cal V} =\left( \begin{array}{cc}
 S&R\\
 0&\widetilde S \end{array} \right),\nn
  \ee
where S and R are $4\times 4$ matrices which in turn have the
block structure:
\be
  S=\left( \begin{array}{cc}
 s_{ij}&a_i \\
 b_j&s \end{array} \right),\quad
 R=\left( \begin{array}{cc}
 a'_i&r_{ij} \\
 s'&b'_j \end{array} \right),\quad i,j=1,\ldots,3.\nn
 \ee
In what follows we use the symbol $\widehat T$ to denote
transposition with respect to the minor diagonal and the symbol $\
\widetilde{A}\ $ to denote $\widetilde A=-A^{\widehat T}$ for a
degenerate matrix $A$ ($\det A= 0$) and $\widetilde
A=(A^{-1})^{\widehat T}$ for a non-degenerate $A$ ($\det A\neq 0$).
In particular,
\be
 (\psi^I)^T=(u^I,v^I),\quad
 \widetilde{\psi}^I=-(v^I,u^I),\quad
 (\widetilde{\psi}^I)^T=-
\left( \begin{array}{cc}
 v^I\\
 u^I \end{array} \right).\nn
 \ee
When applied to a matrix written in the block form, this means:
\be
 \widetilde{S}=\left( \begin{array}{cc}
 s^{-1}&\widetilde{a}_i\\
 \widetilde{b}_j&\widetilde{s}_{ij} \end{array} \right).\nn
 \ee
In the above blocks we use the following $3\times 3$ matrix
potentials:

\be
 \widehat\phi=\left( \begin{array}{ccc}
 \phi_1&0&0\\
 0&\phi_2&0\\
 0&0&\phi_3  \end{array} \right),\quad
  \widehat\chi=\left( \begin{array}{ccc}
 0&0&0\\
 0&0&\chi\\
 0&0&0  \end{array} \right),\nn
 \ee
\be\label{hat_omega_mu}
 \Psi_{3}=\left( \begin{array}{ccc}0&u^3&-v^3\\
 0&0&0\\
 0&0&0  \end{array} \right),\quad
 \widehat\omega=\left( \begin{array}{ccc}\omega_7&\omega_8&0\\
 0&0&0\\
 0&0&0  \end{array} \right),\quad
 \widehat{\mu}_3=\left( \begin{array}{ccc}
 0&0&0\\
 -\mu_3&0&0\\
 0&0&0 \end{array} \right),\nn
 \ee
and the 3-columns:
 \be\label{Psi_a}
 \Psi_a=\left( \begin{array}{ccc}
 u^a\\
 v^a\\
 0 \end{array} \right),\quad
 \widehat{\mu}_a=\left( \begin{array}{ccc}
 \mu_a\\
 0\\
 0 \end{array} \right).\nn
  \ee
  An explicit
exponentiation in (\ref{def_Nu}) gives the following $4\times 4$
blocks for the partial coset representatives:
\be
 {\cal V}_H:\quad S_H=
 \left( \begin{array}{cc}
 \e^{\frac{1}{\sqrt{2}}\widehat\phi}&0\\
 0&\e^{\frac{1}{\sqrt2}\phi_4}\end{array} \right),\quad
  R_H\equiv 0,\nn
 \ee
\be
 {\cal V}_X:\quad S_X=
 \left( \begin{array}{cc}
 \e^{-\widehat\chi}&0\\
 0&1 \end{array}\right),\quad
 R_X\equiv 0,\nn
  \ee
\be
  {\cal V}_{\Psi}:\quad
 S_{\Psi}=
 \left( \begin{array}{cc}
 \e^{\Psi_{3}}&\e^{\frac12\Psi_{3}}\widetilde{\Psi}_1^T\\
 0&1\\
 \end{array} \right),\quad
  R_{\Psi}=
 \left( \begin{array}{cc}
 \e^{\frac12\Psi_{3}}\widetilde{\Psi}_2^T&\Psi\\
 0&\Psi_2^T \e^{\frac12\widetilde{\Psi}_{3}}\\
 \end{array} \right),\nn
 \ee
where
 \be
 \Psi= \frac12\left(\Psi_{12}
 +\frac13(\Psi_{3}\Psi_{12}
+\Psi_{12}\widetilde{\Psi}_{3})
 +\frac{1}{12}
 \Psi_{3}\Psi_{12}\widetilde{\Psi}_{3}\right)+\frac12(1\leftrightarrow 2).\nn
  \ee
 The remaining exponentials are:
\be
 {\cal V}_{\Omega}:\quad S_{\Omega}=\left( \begin{array}{cc}
 1&0\\
 0&1\\
 \end{array} \right),\quad
 R_{\Omega}=
 \left( \begin{array}{cc}
 0&\Omega\\
 0&0\\
 \end{array} \right),\quad
 \Omega=\widehat\omega+\widetilde{\widehat\omega}.\nn
  \ee
\be
 {\cal V}_P:\quad S_P=
 \left( \begin{array}{cc}
 1&\widehat{\mu}_2\\
 0&1\\
 \end{array} \right),\quad
 R_P=
\left( \begin{array}{cc}
 \widehat{\mu}_1&-\Pi-\widehat{\mu}_2\otimes\widetilde{\widehat{\mu}}_1\\
 0&\widetilde{\widehat{\mu}}_1\\
 \end{array} \right),\quad
 \Pi=\widehat{\mu}_3+\widetilde{\widehat{\mu}}_3.\nn
  \ee
For the $K$-involution matrix we have the following block
representation:
\be
 S_K={\cal E},\quad {\cal E}=
 \left( \begin {array}{cc}
 \widehat\kappa&0\\
 0&1
 \end {array} \right), \quad \widehat\kappa={\rm diag}(\kappa,\kappa,1),
 \quad
  \widetilde{\cal E}=\kappa{\cal E},\quad
 R_K\equiv 0.\nn
 \ee
Multiplying all  the matrices ${\cal V}_H,{\cal V}_X,{\cal
V}_{\Psi},{\cal V}_{\Omega}$ and ${\cal V}_{P}$ we obtain for the
coset representative ${\cal V}$:
\be
 {\cal V}=\left( \begin {array}{cc}
 S&R\\
 0&\widetilde{S}
  \end {array} \right),\quad \widetilde{S}=(S^{-1})^{\widehat{T}},\nn
 \ee
where
\be
 S=\left( \begin {array}{cc}
 \e^{\frac{1}{\sqrt2}\widehat\phi}\e^{-\widehat\chi}e^{\Psi_{3}},&
 \e^{\frac{1}{\sqrt2}\phi}e^{-\widehat\chi}\widehat{\rho}_{21}\\
 0&\e^{\frac{1}{\sqrt2}\phi_4}\\
 \end {array} \right),\quad
 \widehat{\rho}_{12}=\widehat{\mu}_1+\e^{\frac12\Psi_{3}}\widetilde{\Psi}_2^T,\quad
 \widehat{\rho}_{21}=\widehat{\mu}_2+\e^{\frac12\Psi_{3}}\widetilde{\Psi}_1^T,\nn
 \ee
\be
 R=\left( \begin {array}{cc}
 \e^{\frac{1}{\sqrt2}\widehat\phi}\e^{-\widehat\chi}\widehat{\rho}_{12},&
 \e^{\frac{1}{\sqrt2}\widehat\phi}\e^{-\widehat\chi}\left(
 \e^{\Psi_{3}}(\Omega-\Pi-\widehat{\mu}_2\otimes\widetilde{\widehat{\mu}}_1)+
 \e^{\frac12\Psi_{3}}\sum_{a=1}^2\widetilde{\Psi}_a^T\otimes
 \widetilde{\widehat{\mu}}_a+\Psi
\right)\\
  0&\e^{\frac{1}{\sqrt2}\phi_4}\widetilde{ \widehat{\rho}}_{12}
 \end {array} \right).\nn
 \ee
Finally, we construct the gauge-invariant representative of the
coset (\ref{def_M}):
\be
 {\cal M}=\left( \begin {array}{cc}
 {\cal P}&{\cal P}{\cal Q}\\
 {\cal Q}^T{\cal P}&\widetilde{\cal P}+{\cal Q}^T{\cal P}{\cal Q}
 \end {array} \right),\quad {\cal Q}=S^{-1}R,\quad
 {\cal P}=S^T{\cal E}S,\quad \widetilde{\cal
P}=\widetilde{S}^T\widetilde{\cal
 E}\widetilde{S},\nn
 \ee
with the block components ${\cal Q}=-{\cal Q}^{\widehat T}$:
\be
 {\cal Q}=\left( \begin {array}{cc}
 \widehat{\mu}_1+\e^{-\frac12\Psi_{3}}\widetilde{\Psi}_2^T,&-\Pi+
 \frac12(\Psi_{21}-\Psi_{12})
 +\Omega-\widehat{\mu}_2\otimes\Psi_2^T +\frac13\Psi_{3}(\frac12\Psi_{12}-
 \Psi_{21})-\widetilde{\Psi}_2^T\otimes\widetilde{\widehat{\mu}}_2
 -\frac13(\frac12\Psi_{12}-
 \Psi_{21})\widetilde{\Psi}_{3}
 \\
  0,&\widetilde{\widehat{\mu}}_1+\Psi_2^T \e^{-\frac12\widetilde{\Psi}_{3}}
 \end {array} \right). \label{def_Q}
 \ee
and ${\cal P}$:
\be
 {\cal P}={\cal P}^T=\left( \begin {array}{cc}
 \e^{\Psi_3^T}\Lambda \e^{\Psi_3},&\e^{\Psi_3^T}\Lambda\widehat{\rho}_{21}\\
 \widehat{\rho}_{21}^T\Lambda
 \e^{\Psi_3},&\widehat{\rho}_{21}^T\Lambda\widehat{\rho}_{21}+\e^{\sqrt2\phi_4}
 \end {array} \right).\label{def_P}
 \ee
The matrix $\Lambda$ entering this expression reads:
\bea
 \Lambda&=&\e^{-\widehat\chi^T}\e^{\frac{1}{\sqrt2}\widehat\phi}\,\widehat\kappa\,
  \e^{\frac{1}{\sqrt2}\widehat\phi}\, \e^{-\widehat\chi}
 = \left( \begin {array}{cc}
 \kappa \e^{\sqrt2\phi_1}&0\\
 0&\widetilde\lambda^0
 \end {array} \right),\nn
  \eea
where the $2\times 2$ matrix $\lambda^0$,  related to $\lambda$ as
$\lambda^0=(X^3)^{-1}\lambda$,  is given by
\be
 \lambda^0=\e^{-\sqrt2\phi_3}\left( \begin {array}{cc}
 1&\chi\\
 \chi&\chi^2+\kappa \e^{\sqrt2(\phi_3-\phi_2)}
 \end {array} \right),\quad
 \widetilde\lambda^0=\kappa \e^{\sqrt2\phi_2}\left( \begin {array}{cc}
 1&-\chi\\
 -\chi&\chi^2+\kappa \e^{\sqrt2(\phi_3-\phi_2)}
 \end {array} \right).\label{def_m_lambda}
 \ee
The following relations are useful:
\be
 \widetilde\lambda_{77}^0=-\frac{\lambda_{77}^0}{\tau_0},\quad
 \widetilde\lambda_{78}^0=\frac{\lambda_{78}^0}{\tau_0},\quad
 \widetilde\lambda_{88}^0=-\frac{\lambda_{88}^0}{\tau_0},\quad
 \tau_0=-\det(\lambda^0)=-\kappa \e^{-\sqrt{2}(\phi_2+\phi_3)},\quad
 \widetilde\tau_0=\tau_0^{-1}.\nn
 \ee

\section{Isometries of the target space}
\subsection{Transformation of the coset}
To classify 28 isometry transformations $SO(4,4)$ of the target
space we consider the action of one-parameter subgroups generated by
\be
 \vec H,\ P^{\pm I},\ W_{\pm I},\ Z_{\pm I},\ \Omega_{\pm p},\ X^{\pm}.
 \label{generators}
\ee
In terms of the gauge-independent coset matrix $\cal{M}$ the
isometries are represented by  (\ref{act_G_on _M}). In conformity
with the matrix representation (\ref{def_matr_basis}) we can
distinguish three  types of the $SO(4,4)$ matrices $g$:
\begin{itemize}
\item The 'right' upper-triangular  matrices generated by the
$B$-type elements of $so(4,4)$:
\be
 g_R=\e^{\alpha\, {\cal C}}=\left( \begin {array}{cc}
 1&\e^{\alpha B}\\
 0&1
 \end {array} \right),\quad {\cal C}=P^1,P^3,W_2,Z_2,\Omega^7,\Omega^8,\nn
 \ee
whose action on the coset components ${\cal P}$ and ${\cal Q}$
consists in the shift
\be\label{PQ_gR}
 {\cal P} \longrightarrow {\cal P}'={\cal P},\quad
 {\cal Q} \longrightarrow {\cal Q}'={\cal Q}+\alpha B.
  \ee
  These correspond to gauge transformations.
\item The ``central'' block-diagonal matrices with the upper-triangular blocks
\be
 g_{Su}=\e^{\alpha\, {\cal C}}=\left( \begin {array}{cc}
 \e^{\alpha A}&0\\
 0&\e^{\alpha\widetilde{A}}
 \end {array} \right),\quad {\cal C}=P^2,W_1,W_3,Z_1,Z_3,X,\nn
 \ee
with the lower-triangular blocks
\be
 g_{Sd}=\e^{\alpha\, {\cal C}}=\left( \begin {array}{cc}
 \e^{\alpha A^T}&0\\
 0&\e^{\alpha\widetilde{A}^T}
 \end {array} \right),\quad {\cal
 C}=P^{-2},W_{-1},W_{-3},Z_{-1},Z_{-3},X^-,\nn
 \ee
and with the diagonal blocks
\be
 g_{S}=\e^{\alpha\, {\cal C}}=\left( \begin {array}{cc}
 \e^{\alpha A_{H_i}^T}&0\\
 0&\e^{\alpha\widetilde{A}_{H_i}^T}
 \end {array} \right),\quad {\cal
 C}=H_i,\quad i=1\ldots 4.\nn
 \ee
These act on the ${\cal P}$ and ${\cal Q}$ blocks  as follows:
\be\label{PQ_gS}
  {\cal P} \longrightarrow {\cal P}'=\e^{\alpha M^T}{\cal P}\e^{\alpha M},\quad
 {\cal Q} \longrightarrow {\cal Q}'=\e^{-\alpha M}{\cal Q}\e^{\alpha
 \widetilde{M}},
  \ee
where $M=A,A^T$ or $A_{H_i}$ for $g=g_{Su},g_{Sd}$ or $g_S$
respectively.
 \item The ``left'' lower-triangular type matrices
\be
 g_L=\left( \begin {array}{cc}
 1&0\\
 \e^{\alpha B^T}&1
 \end {array} \right),\nn
 \ee
whose action on the ${\cal P}$ and ${\cal Q}$ is highly non-trivial
\bea
 {\cal P}'&=&{\cal P}+\alpha{\cal P}{\cal Q}B^T+\alpha B{\cal Q}^T{\cal P}
 +\alpha^2 B(\widetilde{\cal P}+{\cal Q}^T{\cal P}{\cal Q})B^T,\nn\\
 {\cal P}'{\cal Q}'&=&{\cal P}{\cal Q}+\alpha B(\widetilde{\cal P}+
 {\cal Q}^T{\cal P}{\cal Q}),\nn\\
 &&\widetilde{\cal P}+{\cal Q}^T{\cal P}{\cal Q}={\rm inv}.\nn
  \eea
  It is this part of isometries which contains the charging
  transformations.
  \end{itemize}
\indent Meanwhile, there exists a reflection symmetry of the root
diagram interchanging positive and negative roots. This enable us to
reparameterize the target space metric  introducing the dual
coordinates   $\Phi_d^A$ in which the positive root generators look
the same as the negative roots generators in terms of the initial
coordinates $\Phi^A$.   Thus the dual coset matrix ${\cal
M}_d(\Phi_d^A)$ constructed from the dual potentials will be
transformed under the action of the lower-triangular generators in
the same way   as the coset matrix ${\cal M}(\Phi^A)$ under the
gauge transformation $g_R$. We find that the dual coset matrix is an
inverse of the initial coset matrix:
\be
 {\cal M}_d=
 \left( \begin {array}{cc}
 {\cal P}_d&{\cal P}_d{\cal Q}_d\\
 {\cal Q}_d^T{\cal P}_d&\widetilde{\cal P}_d+{\cal Q}_d^T{\cal P}_d{\cal Q}_d
 \end {array} \right)=
 {\cal M}^{-1}=\left( \begin {array}{cc}
 {\cal P}^{-1}+{\cal Q}\widetilde{\cal P}^{-1}{\cal Q}^{T}&-{\cal Q}
 \widetilde{\cal P}^{-1}\\
 -\widetilde{\cal P}^{-1}{\cal Q}^T&\widetilde{\cal P}^{-1}
 \end {array} \right).\nn
 \ee
The transformation of the ${\cal M}$ components under $g_L$ is then
described as the shift of  the dual matrix ${\cal Q}_d$
\be
 \widetilde{\cal P}_d^{-1} \longrightarrow (\widetilde{\cal P}_d^{-1})'=
 \widetilde{\cal P}_d^{-1},\quad
 {\cal Q}_d \longrightarrow {\cal Q}'_d={\cal Q}_d-\alpha B.\nn
 \ee
\subsection{Finite transformations explicitly}
From now on  we will assume  $\kappa=-1$. Using Eqs. (\ref{PQ_gR})
and (\ref{PQ_gS}) it easy to find the finite actions of the Cartan
and the positive-root transformations. The diagonal ones give:
\be
 H_1: \quad \phi_1'=\phi_1+2\alpha_1,\quad \omega_p'=\omega_p
 \e^{-\alpha_1},
 \quad \psi_3'=\psi_3 \e^{-\alpha_1},\quad \mu_a'=\mu_a \e^{-\alpha_1},\nn
  \ee
 \bea
 H_2: \quad \phi_2'&=&\phi_2+2\alpha_2,\quad \omega_8'=\omega_8
 \e^{-\alpha_2}, \quad v_1'=v_1 \e^{-\alpha_2},\quad
 v_2'=v_2 \e^{-\alpha_2},\nn\\
 u_3'&=&u_3 \e^{\alpha_2},\quad
  \mu_3'=\mu_3 \e^{-\alpha_2},\quad \chi'=\chi \e^{-\alpha_2},\nn
  \eea
   \bea
 H_3: \quad \phi_3'&=&\phi_3+2\alpha_3,\quad \omega_7'=\omega_7
 \e^{-\alpha_3}, \quad u_1'=u_1 \e^{-\alpha_3},\quad
 u_2'=u_2 \e^{-\alpha_3},\nn\\
 v_3'&=&v_3 \e^{\alpha_3},\quad
  \mu_3'=\mu_3 \e^{-\alpha_3},\quad \chi'=\chi \e^{\alpha_3},\nn
  \eea
\be
 H_4: \quad \phi_4'=\phi_4+2\alpha_4,\quad \psi_1'=\psi_1 \e^{\alpha_4},
 \quad \psi_2'=\psi_2 \e^{-\alpha_4},
 \quad \mu_1'=\mu_1 \e^{-\alpha_4},\quad \mu_2'=\mu_2
 \e^{\alpha_4}.\nn
  \ee
The   upper-triangular matrices (positive-roots) produce gauge
transformations of the potentials $\Phi^A$. Namely, the generators
$P^I$ and $\Omega_p$ give   simple gauge shifts of the potentials
$\mu_I$ and the KK-potentials $\omega_p$ respectively:
\be
 \mu_I'=\mu_I+\alpha_{\mu}^I,\quad
 \omega_p'=\omega_p+\alpha_{\omega}^p,\quad \hbox{other potentials
 invariant}.\nn
 \ee
The generators $W_I$ and $Z_I$ of the electromagnetic sector produce
the gauge shift of the axions $u^I$ and $v^I$ respectively,
 changing also some other potentials:
\bea
 W_I:\qquad (u^I)'&=&u^I+\alpha_{u}^I,\quad \mu_J'=\mu_J+\frac12
 \delta_{IJK}\alpha_{u}^I
 v^K, \quad \hbox{no sum over }I,K\nn\\
 \omega_7'&=&\omega_7+\alpha_{u}^I
 \mu_I+\frac16\sum_{J,K}\delta_{IJK}\alpha_{u}^I u^J v^K,\quad
 \omega_8'=\omega_8+\frac16\sum_{J,K}\delta_{IJK}\alpha_{u}^I v^J
 v^K,\quad \hbox{no sum over} I.\nn
  \eea
  \bea
 Z_I:\qquad (v^I)'&=&v^I+\alpha_{v}^I,\quad \mu_J'=\mu_J-\frac12
 \delta_{IJK}\alpha_{v}^I
 u^K, \quad \hbox{no sum over }I,K\nn\\
 \omega_8'&=&\omega_8+\alpha_{v}^I
 \mu_I-\frac16\sum_{J,K}\delta_{IJK}\alpha_{v}^I u^J v^K,\quad
 \omega_7'=\omega_7-\frac16\sum_{J,K}\delta_{IJK}\alpha_{v}^I u^J
 u^K,\quad \hbox{no sum over }I.\nn
  \eea
Finally, the generators $X$ and $X^{-}$ lead to finite
transformations:
\bea
 X:\quad &&(v^I)'=v^I+\alpha_{\chi}u^I,\quad
 \omega_8'=\omega_8+\alpha_{\chi}\omega_7,\nn\\
 &&\lambda_{77}^{0'}=\lambda_{77}^0,\quad
 \lambda_{78}^{0'}=\lambda_{78}^0+\alpha_{\chi}\lambda_{77}^0,\quad
 \lambda_{88}^{0'}=\lambda_{88}^0+2\alpha_{\chi}\lambda_{78}^0+
 \alpha_{\chi}^2\lambda_{77}^0\nn\\
 X^{-}:\quad &&(u^I)'=u^I+\alpha_{-\chi}v^I,\quad
 \omega_7'=\omega_7+\alpha_{-\chi}\omega_8,\nn\\
 &&\lambda_{88}^{0'}=\lambda_{88}^0,\quad
 \lambda_{78}^{0'}=\lambda_{78}^0+\alpha_{-\chi}\lambda_{88}^0,\quad
 \lambda_{77}^{0'}=\lambda_{77}^0+2\alpha_{-\chi}\lambda_{78}^0+
 \alpha_{-\chi}^2\lambda_{88}^0,\nn
 \eea
with the remaining potentials invariant.
\subsection{Killing vectors}
 To find the differential operators generating isometries (the Killing vectors
 $X_k$ of the target space) from the finite transformations one
can use the defining equations
 \be
  X_k=\frac{\partial\Phi^{A'}}{\partial\alpha_k}\Big|_{\alpha_k=0}\,
  \frac{\partial}{\partial\Phi^A},\nn
  \ee
where $\Phi^{A'}=\Phi^{A'}(\Phi^B,\alpha_k)$ are the potentials
transformed under the action of the one-parametric subgroups. Here
we give $X_k$ corresponding to the Cartan, positive-root and $X^{-}$
generators. The others Killing vectors are  much more complicated
and we will present them in the next section only for the vacuum
seed potentials. Enumerating the potentials as
$\Phi^A=(X^1,X^2,\lambda_{pq}^0,\psi^I,\mu_I,\omega_p)$ and the
parameters as
$\alpha_k=(\alpha_1,...,\alpha_4,\alpha_{u}^I,\alpha_{v}^I,
  \alpha_{\mu}^I,\alpha_{\omega}^p,\alpha_{\chi},\alpha_{-\chi})$
 we find the following set of the Killing vectors:
 \bea
  &&M_p{}^q=2\lambda_{pr}^0\frac{\partial}{\partial\lambda_{rq}^0}+
  \omega_p\frac{\partial}{\partial\omega_q}+\delta_{p}^q\,\omega_r
  \frac{\partial}{\partial\omega_r}+\delta_p^q\,\mu_I
  \frac{\partial}{\partial\mu_I}
  +\psi_p^I\frac{\partial}{\partial\psi_q^I},\nn\\
  &&\frac{H_1+H_4}{\sqrt2}=2(X^1)^2X^2\frac{\partial}{\partial X^1}
  +\psi_p^1\frac{\partial}{\partial \psi_p^1}
  -\psi_p^2\frac{\partial}{\partial \psi_p^2}-\psi_p^3
  \frac{\partial}{\partial \psi_p^3}
  -2\mu_1\frac{\partial}{\partial
  \mu_1}-\omega_p\frac{\partial}{\partial\omega_p},\nn\\
 &&\frac{H_1-H_4}{\sqrt2}=2(X^2)^2X^1\frac{\partial}{\partial X^2}
  +\psi_p^2\frac{\partial}{\partial \psi_p^2}
  -\psi_p^1\frac{\partial}{\partial \psi_p^1}-\psi_p^3
  \frac{\partial}{\partial \psi_p^3}
  -2\mu_2\frac{\partial}{\partial
  \mu_2}-\omega_p\frac{\partial}{\partial\omega_p},\nn\\
  &&\Sigma_I^p=\frac{\partial}{\partial\psi_p^I}+\frac12\delta_{IJK}
  \epsilon^{pq}\psi_q^K\frac{\partial}{\partial
  \mu_J}+\mu_I\frac{\partial}{\partial
  \omega_p}+\frac16\delta_{IJK}\epsilon^{pq}\psi_q^J\psi_r^K
  \frac{\partial}{\partial
  \omega_r},\nn\\
  &&\Omega^p=\frac{\partial}{\partial\omega_p},\quad P^I=
  \frac{\partial}{\partial
  \mu_I},\nn
  \eea
 where $M_p{}^q$ are given by
 \be
  M_7{}^7=-\frac{1}{\sqrt2}(H_1+H_3),\quad M_8{}^8=-\frac{1}{\sqrt2}(H_1+H_2),\quad
  M_7{}^8=X,\quad M_8{}^7=X^{-}.\nn
  \ee

\section{Solution generating technique}
The use of the target space isometries for generating purposes
consists in three steps. First, one has to choose the seed solution
and to find the corresponding target space potentials. This involves
solving the (differential) dualisation equations. Then the isometry
transformations are applied to get the target space potentials of
the new solution. Finally one has to solve back the dualisation
equations (\ref{dua2}) to obtain new solution in terms of the metric
and the matter fields. The three-dimensional metric $h_{ij}$ remains
essentially the same.

 Since the dimensional reduction from eleven to five
dimensions does not involve dualisation, an identification of
solution in five-dimensional or in eleven-dimensional terms is the
matter of choice. Five target space variables
$\phi_1,\,\phi_2,\,\phi_3,\,\phi_4,\, \chi$ enter the
eleven-dimensional metric algebraically, via the moduli $X^I,\,
\lambda_{pq}$: \be
  ds_{11}^2 = \sum_{I,i,i'} X^I \left( (dz^i)^2 + (dz^{i'})^2
  \right)+
  \lambda_{pq}(dz^p+a^p)(dz^q+a^q)+\tau^{-1}h_{ij}dx^idx^j,\nn
  \ee
while the KK vectors $a^p$ in the $T^2$ sector are related to the
target space potentials $\omega_p$ via dualisation. In the
form-field sector, \be
  A_{[3]}=(A^1+\psi_p^1 dz^p)\wedge dz^1 \wedge dz^2+
 (A^2+\psi_p^2 dz^p)\wedge dz^3 \wedge dz^4+
 (A^3+\psi_p^3 dz^p)\wedge dz^5 \wedge dz^6.\nn
  \ee
  the six quantities $\psi_p^I$ are the target space
  potentials, while the remaining one forms $A^I$ are related to the
  potentials $\mu_I$ via dualisation.
\subsection{Asymptotic conditions}
To find a proper direction in the target space which would lead to
the solution with desired properties is a non-trivial task, and
usually it invokes an identification of the subgroups of the
isometry group preserving certain asymptotic conditions for the
metric (and/or the form field). For the black hole/black ring
applications several such conditions are of interest.
\begin{itemize} \item{\bf Minkowskian metric}

Consider the eleven-dimensional Minkowski metric in the Cartesian
coordinates (assuming $\kappa=-1$)
 \be
  ds_{11}^2=\sum_{k=1}^7 (dz^k)^2-(dz_8)^2+\sum_{i=1}^3 (dx^i)^2,\quad
  A_{[3]}=0.\label{asym_cartesian}
  \ee
This  correspond to $\lambda_{88}=-1,\lambda_{77}=1$ and all other
potentials zero. Consequently, the coset matrix  ${\cal M}_{as}=K$.
By virtue of Eq.(\ref{h_K_h}), such an asymptotic  is preserved
under isometries belonging to the isotropy subgroup $H$ of the
$SO(4,4)$:
\be
 P^I+ P^{-I},\quad Z_I+ Z_{-I},\quad W_I-W_{-I},\quad X+
 X^{-},\quad \Omega^7+ \Omega^{-7},\quad
 \Omega^8-\Omega^{-8}.\nn
 \ee
\item{\bf Flat metric in the $ S^2\times S^1$ fibration}

Another useful form of the flat metric is appropriate to the
five-dimensional ring problems is
 \be
  ds_5^2=-(dt)^2+r^2\cos^2\theta(d\psi)^2+dr^2+r^2(d\theta^2+\sin^2\theta
  d\phi^2),
  \nn
  \ee
  and the reduction is performed along $t,\,\psi$ where $\psi$ have the sense
 of the angular variable along $S^1$ in the ring $ S^2\times S^1$ fibration.
 We identify
  $z^7=\psi, z^8=t$, then the target space variables
$\lambda_{88}=-1,\lambda_{77}=\tau=r^2\cos^2\theta$. Preservation of
the above line element is more restrictive: from an analysis of an
infinitesimal action of generators (\ref{generators}) we  find the
 only combination  of the  Killing vectors  \be Z_I+Z_{-I}.\nn\ee
Mathematically, this is related to the coordinate dependence of the
asymptotical matrix ${\cal M_{as}}$.
\item{{\bf Guisto-Saxena coordinates}}

However one can perform dimensional reduction with respect to the
combinations $\phi_{\pm}=\frac12(\phi \pm \psi)$ instead of $\psi$,
as suggested in the Ref. \cite{gisa}. In this case the coset matrix
${\cal M_{as}}$ will be coordinate independent. The target space
potentials then read
 \be
  \lambda_{77}=\tau,\quad \lambda_{88}=-1,\quad
  \omega_7=\tau,\quad \tau=r^2 \label{GS_cond}
  \ee
(other potentials zero), so the coset matrix ${\cal{M}}_{as}$ will
be given through
 the following constant $4\times 4$ blocks:
\be
 {\cal{P}}_{as}=\left( \begin {array}{cccc}
 -\tau^{-1}&0&0&0\\
 0&-1&0&0\\
 0&0&\tau^{-1}&0
 \\0&0&0&1\end {array} \right),\quad
 {\cal{Q}}_{as}=\left( \begin {array}{cccc}
 0&\tau&0&0\\
 0&0&0&0\\
 0&0&0&-\tau
 \\0&0&0&0\end {array} \right).\nn
 \ee
To find the appropriate combinations of the Killing vectors
consider their vacuum form setting all  electromagnetic potentials
to zero and taking $X^I=1$ (in this case $\lambda^0=\lambda$):
 \bea
 &&M_p{}^q=2\lambda_{pr}\frac{\partial}{\partial\lambda_{rq}}+
  \omega_p\frac{\partial}{\partial\omega_q}+\delta_{p}^q\,\omega_r
  \frac{\partial}{\partial\omega_r},\nn\\
  &&\frac{H_1+H_4}{\sqrt2}=2\frac{\partial}{\partial X^1}
  -\omega_p\frac{\partial}{\partial\omega_p},\quad
  \frac{H_1-H_4}{\sqrt2}=2\frac{\partial}{\partial X^2}
  -\omega_p\frac{\partial}{\partial\omega_p},\nn\\
  &&\Sigma_I^p=\frac{\partial}{\partial\psi_p^I},\quad
  \Sigma_{-I}^p=\omega_p\frac{\partial}{\partial\mu_I}+\lambda_{pq}
  \frac{\partial}{\partial\psi_q^I},\quad
   P^{I}=\frac{\partial}{\partial\mu_I},\quad
   P^{-I}=-\omega_p\frac{\partial}{\partial\psi_p^I},\nn\\
   &&\Omega^p=\frac{\partial}{\partial\omega_p},\quad
   \Omega^{-p}=-2\lambda_{pr}\omega_s\frac{\partial}{\partial\lambda_{rs}}-
   (\omega_p\omega_r+\tau\lambda_{pr})\frac{\partial}{\partial\omega_r}.\nn
  \eea
Conditions (\ref{GS_cond}) then give the following linear
combinations preserving ${\cal{M}}_{as}$:
\be
 Z_I+Z_{-I},\quad W_{-I}+P^{-I},\quad X-\Omega^{-8}.\nn
 \ee
\end{itemize}
More general physically interesting asymptotic conditions may be
encountered for five-dimensional type D metrics as discussed in
\cite{Pravda:2007ty}.
\section{Three-charge black hole with two angular momenta}
Five-dimensional stationary charged black holes were investigated
for different couplings of vector fields to gravity both for
non-supersymmetric \cite{ Cvetic:1996xz, Ida:2003wv,clp2004,
Morisawa:2004tc, Kunz:2005ei, Aliev:2006, Kunz:2005nm, Kunz:2006xk,
Kunz:2006jd,Kunz:2006yp, Kunz:2006eh, Mei:2007bn, Kleihaus:2007kc}
and supersymmetric \cite{Cacciatori:2004qm, Bellorin:2006yr}
configurations (for a recent review see \cite{Emparan:2008eg}).
Somewhat surprisingly, the simplest Einstein-Maxwell theory in five
dimensions does not possess analytic solutions like the Kerr-Newman
black hole in four dimensions. This is related to the lack of hidden
symmetries which are enhanced in the supergravity action due to
Chern-Simons term. The enhancement endows us with charging
symmetries which open the way  to construct the three-charge doubly
rotating black hole solution from the $5D$ vacuum Myers-Perry
metric. We assume the following choice of coordinates:
 $z_7=\psi,\,z_8=t,\ r,\theta,\phi$, and denote the rotation
  parameters as $a,b$. The seed solution then reads:
\begin{equation}
ds^2 = - dt^2 + \frac{\rho^2r^2}{\Delta} dr^2 + \rho^2 d\theta^2 +
(r^2 + a^2) \sin^2\theta d\phi^2 + (r^2 + b^2) \cos^2\theta d\psi^2
+ p \left( dt + a \sin^2\theta d\phi + b \cos^2\theta d\psi
\right)^2.\label{5DKerr}
\end{equation}
Using the relations between the metric components and the
target-space potentials:
  \bea
 &&g_{tt}=\lambda_{88},\quad g_{t\psi}=\lambda_{78},\quad
 g_{\psi\psi}=\lambda_{77},\nn\\
 && g_{t\phi}=\lambda_{88}a_{\phi}^8+\lambda_{78}a_{\phi}^7,\quad
 g_{\phi\psi}=\lambda_{78}a_{\phi}^8+\lambda_{77}a_{\phi}^7,\nn\\
 &&g_{\phi\phi}=\lambda_{88}(a_{\phi}^8)^2+2\lambda_{78}a_{\phi}^7
 a_{\phi}^8+\lambda_{77}(a_{\phi}^7)^2+\tau^{-1}h_{\phi\phi},\nn
  \eea
we find the $\sigma$-model variables:
\begin{eqnarray}
\lambda_{88} &=& - 1 + p, \quad \lambda_{78} = p b \cos^2\theta,
\quad \lambda_{77} = (r^2 + b^2) \cos^2\theta + p b^2 \cos^4\theta,
\quad \tau = \left( r^2 + b^2 - r_0^2 + p a^2 \cos^2\theta \right)
\cos^2\theta,
\nonumber\\
a^7_\phi &=& \tau^{-1} p a b \sin^2\theta \cos^2\theta,\quad
a^8_\phi = - \tau^{-1} p (r^2 + b^2) a \sin^2\theta \cos^2\theta,
\quad \omega_7=-p a b \cos^4\theta,\quad \omega_8 = - p a
\cos^2\theta.\nn
\end{eqnarray}
The invariant three-metric reads
 \be
  h_{ij} dx^i dx^j = \tau \left( \frac{\rho^2 r^2}{ \Delta} dr^2 + \rho^2
d\theta^2 + \frac{\Delta}{\tau} \sin^2\theta \cos^2\theta d\phi^2
\right), \qquad \sqrt{h} = \frac12 \tau \rho^2 \sin\theta
\cos\theta,\nn
  \ee
where \vspace{-.5cm}
 \be
  p=\frac{r_0^2}{\rho^2},\quad
  \rho^2=r^2+a^2\cos^2\theta+b^2\sin^2\theta,\quad
  \Delta=(r^2+ a^2)(r^2 + b^2) - (r_0 r)^2.\nn
  \ee
To equip this vacuum solution with three electric charges, we
perform the following transformation:
 \be
 {\cal M}'=\Pi^T  {\cal M}
 \Pi,\quad \Pi=\prod_I \e^{\alpha_I(Z_I+Z_{-I})},\nn
 \ee
where the product of one-parametric exponentials reads explicitly:
 \be
 \Pi=
 \left( \begin {array}{cccccccc} c_3&0&-s_3&0&0&0&0&0\\0&c_1c_2&0&-
 s_1c_2&-c_1s_2&0&-s_1s_2&0\\
 -s_3&0&c_3&0&0&0&0&0\\0&-s_1c_2&0&c_1c_2&s_1s_2&0&c_1s_2&0
 \\0&-c_1s_2&0&s_1s_2&c_1c_2&0&s_1c_2&0\\0&0&0&0&0&c_3&0&s_3\\
 0&-s_1s_2&0&c_1s_2&s_1c_2&0&c_1c_2&0\\0&0&0&0&0&s_3&0&c_3\end {array}
 \right),\quad c_I\equiv \cosh(\alpha_I),\quad  s_I\equiv
 \sinh(\alpha_I).\nn
  \ee
Using formulae of the Appendix D, one can extract the transformed
potentials in the following form:
 \bea
 g_{\psi\psi}'=\lambda_{77}'&=&D^{-2/3}\cos^4\theta\left(p\,b^2-
 p^2\Bigl(s^2(a^2+b^2)+2a\,b\,c\,s+a^2\sum_{I<J}
 s_I^2s_J^2\Bigr)-(a^2-b^2)\Bigl(1+p\sum_I s_I^2\Bigr) \right)\nn\\
 &+&D^{1/3}\rho^2\cos^2\theta,\nn\\
 g_{t\psi}'=\lambda_{78}'&=&D^{-2/3}\cos^2\theta p(b\,c+s\,a-p\,s\,a),\nn\\
 g_{tt}'=\lambda_{88}'&=&D^{-2/3}(p-1),\nn\\
 \tau'&=&D^{-1/3}\tau,\quad (X^I)'=\frac{D^{1/3}}{D_I}.\nn\\
  (u^I)'&=&\frac{p\cos^2\theta}{D_I}(a\,c_Is_Js_K+b\,s_Ic_Jc_K),\quad
  (v^I)'=\frac{p\,c_Is_I}{D_I},\nn\\
  (\mu_I)'&=&-\frac{D_J+D_K}{2D_JD_K}p\cos^2\theta(a\,s_Ic_Jc_K+b\,c_Is_Js_K),\quad
  I\neq J\neq K.\nn
  \eea
   \bea
   \omega_7'&=&\omega_7-\frac{p^2\cos^4\theta}{D}\Biggl\{cs(a^2+b^2)+2abs^2+
   ab\sum_{I<J} s_I^2s_J^2 +\frac13p\Bigl(cs(a^2+b^2)\sum_I s_I^2\nn
   \\ &+&ab\Bigl(\sum_{I<J} s_I^2s_J^2(s_I^2+s_J^2)-\sum_{I<J}
   s_I^2s_J^2+3s^2+2s^2\sum_I
   s_I^2\Bigr)\Bigr)+p^2abs^2\Biggr\},\nn\\
   \omega_8'&=&c\omega_8+\frac{p\cos^2\theta}{D}\Biggl\{bs+psb\Bigl(3+
   \sum_I s_I^2\Bigr)+\frac13p^2\Bigl((ca+sb)\sum_{I<J} s_I^2s_J^2+3sb\sum_I
   s_I^2\Bigr)+p^3acs^2\Biggr\},\nn
  \eea
where
 \be
  c\equiv \prod_I c_I,\quad s\equiv \prod_I s_I,\quad
  D\equiv \prod_I D_I,\quad D_I=1+ps_I^2.\nn
  \ee
Note that metric (\ref{5DKerr}) is invariant under an interchange of
the angular coordinates $\psi$ and $\phi$ together with the
corresponding rotation parameters
 $\phi\leftrightarrow \psi,\; \theta \leftrightarrow
  \theta+\pi/2,\; a \leftrightarrow b$, namely,
 $g_{tt}\leftrightarrow g_{tt},\; g_{t\psi}\leftrightarrow
  g_{t\phi},\;
  g_{\psi\psi}\leftrightarrow g_{\phi\phi},\; g_{\psi\phi}\leftrightarrow
  g_{\psi\phi},\;
  A_{\psi}^I\leftrightarrow A_{\phi}^I.$
We assume that this symmetry remains valid for the charged solution
(\ref{5DKerr}) as well. This simplifies an extraction of the final
form of the solution avoiding the inverse dualisation. As a result
we obtain  the five-dimensional one-forms $A^I$ and the metric
components as follows:
 \bea
  A^I&=&\frac{p}{D_I}\Bigl(s_Ic_Idt+(b\,c_Is_Js_K+a\,s_Ic_Jc_K)\sin^2\theta d\phi+
  (a\,c_Is_Js_K+b\,s_Ic_Jc_K) \cos^2\theta d\psi\Bigr),\ I\neq J\neq K\nn\\
  g_{\phi\phi}'&=&D^{-2/3}\sin^4\theta\left(p\,a^2-p^2\Bigl(s^2(a^2+b^2)+2a\,b\,c\,s+b^2\sum_{I<J}
 s_I^2s_J^2\Bigr)+(a^2-b^2)\Bigl(1+p\sum s_I^2\Bigr) \right)\nn\\
 &+&D^{1/3}\rho^2\sin^2\theta,\nn\\
 g_{t\phi}'&=&D^{-2/3}\sin^2\theta p(a\,c+s\,b-p\,s\,b),\nn\\
 g_{\psi\phi}'&=&D^{-2/3}\cos^2\theta\sin^2\theta p\left(ab-
 p\Bigl(ab\sum_{I<J} s_I^2s_J^2+(a^2+b^2)cs+2abs^2\Bigr)\right),\nn\\
 h_{ij}'&=&h_{ij},\quad
 g_{rr}'=D^{1/3}\frac{\rho^2r^2}{\Delta},\quad
 g_{\theta\theta}'=D^{1/3}\rho^2.\nn
  \eea
This solution generalizes the solution found by Cvetic, Lu and Pope
\cite{clp2004} for equal rotation parameters within the gauged $5D$
supergravity, reducing to the latter for the gauge-coupling constant
$g=0$, an identification $a=b=l$ under relabeling
 $s_I \rightarrow -s_I,\; \psi,\phi
\rightarrow -\psi,-\phi,\; A^I\rightarrow -A^I$. The three-charge
doubly rotating black hole solution within the compactified
heterotic theory was given in \cite{Cvetic:1996xz}.

\section{$G_{2(2)}$ embedded in $SO(4,4)$}
The present model reduces to minimal five-dimensional supergravity
under the following indentifications\be
\psi^1=\psi^2=\psi^3=\psi,\quad \mu_1=\mu_2=\mu_3=\mu,\quad
\lambda^0=\lambda, \quad X^1=X^2=X^3=1  \label{cond_on_pot},\ee
leading to the relations:
  \be \phi_1=\frac{1}{\sqrt2}(\varphi_2+\frac{1}{\sqrt3}\varphi_1)
 ,\quad \phi_2=\frac{1}{\sqrt2}(\varphi_2-\frac{1}{\sqrt3}\varphi_1),
 \quad \phi_3=\sqrt{\frac23}\varphi_1,\quad
 \phi_4=0. \nn \ee
 In this case the target space metric of the three-dimensional
sigma-model will read:
\begin{eqnarray}\label{g2contracted}
dl^2&=& \frac12 \big(d\varphi_1^{2}+d\varphi_2^{2}+3\kappa
 \e^{\varphi_2+\frac{1}{ \sqrt3}\varphi_1} G^2 +
 3\e^{\frac{2}{\sqrt3}\varphi_1} du^{2} +3\kappa
 \e^{\varphi_2-\frac{1}{\sqrt3}\varphi_1} (dv -\chi du)^{2}+\kappa
 \e^{\varphi_2+\sqrt3\varphi_1} G_{7}^{2}
\nn \\ &+&e^{2\varphi_2}G_{8}^2+\kappa
 \e^{\varphi_2-\sqrt3\varphi_1}d\chi^{2} \big),\end{eqnarray} where
the one-forms are: \vspace{-.4cm}\begin{eqnarray}
G&=&d\mu+vdu-udv,\nn
\\
 G_{7}&=&V_7,\quad G_{8}=V_8-\chi V_7,\nn\\
 V_p&=&d\omega_p-\psi_p\left(3d\mu+\epsilon^{qt}d\psi_q
\psi_t\right).\nn
\end{eqnarray}
This manifold is invariant under the $G_{2(2)}$ subgroup of the
$SO(4,4)$. Dimensional reduction of the $5D$ minimal supergravity to
three dimensions was recently studied in \cite{bccgsw,gc07}. In the
notation of \cite{bccgsw,gc07} the indices $p,q,t$ take the values
$0,1$, the coordinate $z^7$ is time-like and the matrix $\lambda$ is
related to the present one by transposition with respect to the
minor diagonal. In matrix terms the target-space metric
(\ref{g2contracted}) reads:
\be
 dl^2 =
\frac14 \mathrm{Tr} \left( \lambda^{-1} d\lambda \lambda^{-1}
d\lambda \right) + \frac14 \tau^{-2} d\tau^2 +\frac32 d\psi^T
\lambda^{-1} d\psi - \frac12\tau^{-1} V^T \lambda^{-1} V -
 \frac32\tau^{-1} G^2.\nn
  \ee
This coincides with the result of \cite{bccgsw,gc07} for the
Euclidean signature of the three-space $(\kappa=-1)$.
\subsection{$g_{2(2)}$ subalgebra of $so(4,4)$}
The above contraction to $G_{2(2)}$ can be described in terms of
the root space as follows. Consider the root vectors of the
$so(4,4)$ algebra in the following basis:
\bea
 \vec {e}_{1}&=&\left(s-\frac12,s+\frac12,\frac12,s\right),\qquad
 \vec{e}_{2}=\left(1,-1,\frac12,s\right),\quad\ \
 \vec{e}_{3}=\left(-s-\frac12,\frac12-s,\frac12,s\right),\nn
 \\
 \vec{e}_{4}&=&\left(\frac12-s,-s-\frac12,1,0\right),\quad\ \
 \vec{e}_{5}=(-1,1,1,0),\quad\
 \
 \vec{e}_{6}=\left(s+\frac12,s-\frac12,1,0\right),\ \nn
 \\
 \vec{e}_{7}&=&\left(\frac12-s,-s-\frac12,-\frac12,s\right),\
 \
 \vec{e}_{8}=\left(-1,1,-\frac12,s\right),\
 \
 \vec{e}_{9}=\left(s+\frac12,s-\frac12,-\frac12,s\right),\ \nn
 \\
 \vec{e}_{10}&=&\left(0,0,\frac32,s\right),\qquad\qquad\quad\ \
 \vec{e}_{11}=(0,0,0,2s),\quad\ \
 \vec{e}_{12}=\left(0,0,-\frac32,s\right),\quad s=\frac{\sqrt3}{2}. \nn
 \eea
Examination of this pattern  shows that the following combinations
of the triplets of the $ so(4,4)$  root vectors
\be
 \vec\alpha_{\pm4}=\frac13\sum_I \vec e_{\pm I},\quad
 \vec\alpha_{\pm1}=\frac13\sum_I \vec e_{\pm
 (I+3)},\quad \vec\alpha_{\pm3}=\frac13\sum_I \vec e_{\pm (I+6)},\nn
 \ee
together with \vspace{-.4cm}
\be
 \vec\alpha_{\pm5}=\vec e_{\pm10},\quad \vec\alpha_{\pm6}=\vec
 e_{\pm11},\quad \vec\alpha_{\pm2}=\vec e_{\pm12},\nn
 \ee
form the standard set of the $g_{2}$ roots  satisfying the
relations:
\be
 \vec\alpha_{\pm3}=\pm(\vec\alpha_{1}+\vec\alpha_{2}),\quad
 \vec\alpha_{\pm4}=\pm(2\vec\alpha_{1}+\vec\alpha_{2}),\quad
 \vec\alpha_{\pm5}=\pm(3\vec\alpha_{1}+\vec\alpha_{2}),\quad
 \vec\alpha_{\pm6}=\pm(3\vec\alpha_{1}+2\vec\alpha_{2}).\nn
 \ee
The corresponding  generators read:
\bea
 && M_1=\frac{\sqrt2}{3}(H_1-H_2+2H_3),\quad M_2=\sqrt{\frac23}
 (H_1+H_2),\nn\\
 &&P^{\pm}=\frac{1}{\sqrt3}\sum P^{\pm I},\quad Z_{\pm}=
 \frac{1}{\sqrt3}\sum Z_{\pm I},
 \quad W_{\pm}=\frac{1}{\sqrt3}\sum W_{\pm I},\quad
 \Omega^{\pm p},\quad X^{\pm}.\nn
 \eea
They obey  the following commutation relations in the Cartan-Weyl
form:
\bea
 && {[}P^{+},P^{-}{]}=\frac12M_1+\frac{\sqrt3}{2}M_2,\nn\\
 && {[}W_{+},W_{-}{]}=M_1,\nn\\
 && {[}Z_{+},Z_{-}{]}=-\frac12M_1+\frac{\sqrt3}{2}M_2,\nn\\
 && {[}\Omega^{7},\Omega^{-7}{]}=\frac32M_1+\frac{\sqrt3}{2}M_2,
 \quad {[}\Omega^{8},\Omega^{-8}{]}=\frac{\sqrt3}{2}M_2,\nn\\
 && {[}X^{+},X^{-}{]}=-\frac32M_1+\frac{\sqrt3}{2}M_2,\nn\\
 && {[}W_{\pm},P^{\pm}{]}=\mp\Omega^{\pm 7},\quad {[}Z_{\pm},P^{\pm}{]}
 =\mp\Omega^{\pm8},\nn\\
 && {[}W_{\pm},Z_{\pm}{]}=\mp P^{\pm},\nn\\
 && {[}X^{\pm},W_{\pm}{]}=\mp Z^{\pm},\nn\\
 && {[}X^{\pm},\Omega^{\pm7}{]}=\mp\Omega^{\pm8},\nn
 \eea
 and so on.

\subsection{$8\times 8$ matrix representation for the coset
$G_{2(2)}/(SL(2,R)\times SL(2,R))$}
 Contracting  the set of the potentials $\Phi^A$  according to the
conditions (\ref{cond_on_pot}), we obtain the following
representation for the coset blocks ${\cal P}$ and ${\cal Q}$:
\be
 {\cal Q}=\left( \begin{array}{ccc}
 \mu,&\omega-\mu\psi^T,&0 \\
 \widetilde{\psi}^T,&\mu\sigma_3,& \widetilde\omega-
 \mu\widetilde{\psi}^T\\
 0,&\psi^T,&-\mu\end{array} \right),\quad \sigma_3=
 \left( \begin{array}{cc}
 1&0 \\
 0&-1\end{array} \right).\nn
 \ee
 \be
 {\cal P}=-\tau^{-1}\left( \begin{array}{ccc}
 1,&\eta^T,&\mu \\
 \eta,&\eta\eta^T-\tau\widetilde{\lambda},&
 \eta\mu-\tau \widetilde{\lambda}\widetilde{\psi}^T \\
 \mu,&
 \eta^T\mu-\tau \widetilde{\psi}\widetilde{\lambda},
 &\mu^2-\tau-\tau\widetilde{\psi}\widetilde{\lambda}\widetilde{\psi}^T
 \end{array} \right),\quad \eta=\sigma_3\psi.\nn
 \ee
This gives a   $8\times 8$ representation of the coset
$G_{2(2)}/(SL(2,R)\times SL(2,R))$  of the minimal
five-dimensional supergravity reduced to three dimensions,
alternative to the $7\times 7$ one given in \cite{bccgsw,gc07}.
\section{Conclusions}
In this paper we have constructed a generating technique for the
$U(1)^3\;  5D$ supergravity with two commuting Killing symmetries.
This theory is reduced to the three-dimensional gravity coupled
sigma model on symmetric spaces $SO(4,4)/SO(4)\times SO(4)$ or
$SO(4,4)/SO(2,2)\times SO(2,2)$ depending on the signature of the
three-space. The classical U-duality group of the three-dimensional
theory is the 28-parametric non-compact group $SO(4,4)$ which acts
transitively on the target space. This enables one to generate new
five-dimensional solutions with the same three-metric from the seed
ones. We were able to obtain finite transformations   in terms of
the target space potentials, and, in addition, we constructed the
$8\times 8$ matrix representation of the coset, which is  convenient
for performing the transformations explicitly. Particular
combinations of  transformations were identified which preserve
asymptotic conditions relevant for the black hole and the black ring
problems. We presented the action of charging transformations on a
neutral seed, assuming the dimensional reduction in terms of
Guisto-Saxena coordinates.

As an application, we have constructed a new rotating
five-dimensional black hole with three independent charges and two
rotation parameters. Our technique allows in principle to generate
black rings with the maximal number of parameters (a mass, two
rotation parameters, three electric charges and three magnetic
dipole moments), but so far our attempts to find such a solution in
a concise form were unsuccessful.

An identification of the three vector fields and freezing out the
two scalar moduli  reduce the present theory  to minimal
five-dimensional supergravity with the three-dimensional U-duality
group $G_{2(2)}$, which was extensively studied recently along
similar lines \cite{bccgsw,gc07}. For this limiting case we have
presented a new matrix representation for the coset
$G_{2(2)}/(SL(2,R)\times SL(2,R))$ in terms of the $8\times 8$
matrices.

\begin{acknowledgments}
The authors thank Gerard Clement and Paul Sorba (LAPTH, Annecy) and
Chiang-Mei Chen (NCU, Taiwan) for helpful  suggestions. The work was
supported by the RFBR grant 08-02-01398-a.

\end{acknowledgments}

\appendix

\section{$8\times 8$ matrix representation }
We choose the following $8\times 8$ matrix representation of the
so(4,4) algebra
 \be
  E= \left( \begin{array}{cc}
A & B \\
C & -A^{\widehat{T}} \end{array} \right),\label{def_matr_basis}
 \ee
 where
$A,\ B,\ C$ are the $4\times 4$ matrices, $A,\ B$ being
antisymmetric, $B=-B^{T},\ C=-C^{T}$, and the symbol $\widehat{T}$
in $A^{\widehat{T}}$ means  transposition with respect to the minor
diagonal. The diagonal matrices $\vec H$ are given by the following
$A-$type matrices (with $B=0=C$): \bea
 A_{H_1}&=&\left( \begin {array}{cccc}
 \sqrt{2}&0&0&0\\0&0&0&0\\0&0&0&0
 \\0&0&0&0\end {array} \right), \;\;
 A_{H_2}=\left(\begin {array}{cccc} 0&0&0&0\\0&\sqrt{2}&0&0\\0&0&0&0
 \\0&0&0&0\end {array} \right),\;\;
 A_{H_3}=\left(\begin {array}{cccc} 0&0&0&0\\0&0&0&0\\0&0&\sqrt{2}&0
 \\0&0&0&0\end {array} \right),\;\;
 A_{H_4}=\left(\begin {array}{cccc} 0&0&0&0\\0&0&0&0\\0&0&0&0
 \\0&0&0&\sqrt{2}\end {array} \right).\nn
 \eea
Twelve generators corresponding to the positive roots are given by
the upper-triangular matrices $E_k,\ k=1,\ldots,12, $.  From these
the generators labeled by $k=2,4,6,7,9,12$ are of pure $A$-type
(with $B=0=C$):
 \bea
 A_{E_2}=\left(\begin {array}{cccc} 0&0&0&1\\0&0&0&0\\0&0&0&0
 \\0&0&0&0\end {array} \right),\quad
 A_{E_4}=\left(\begin {array}{cccc} 0&0&0&0\\0&0&0&0\\0&0&0&-1
 \\0&0&0&0\end {array} \right), \quad
 A_{E_6}=\left(\begin {array}{cccc} 0&1&0&0\\0&0&0&0\\0&0&0&0
 \\0&0&0&0\end {array} \right),\quad \nn\\
 A_{E_7}=\left(\begin {array}{cccc} 0&0&0&0\\0&0&0&-1\\0&0&0&0
 \\0&0&0&0\end {array} \right)\quad
 A_{E_9}=\left(\begin {array}{cccc} 0&0&-1&0\\0&0&0&0\\0&0&0&0
 \\0&0&0&0\end {array} \right),\quad
A_{E_{12}}=\left(\begin {array}{cccc} 0&0&0&0\\0&0&-1&0\\0&0&0&0
 \\0&0&0&0\end {array} \right).\nn\eea
 while the other six are of pure $B$ type (with $A=0=C$):
 \bea B_{E_1}=\left(\begin {array}{cccc} 1&0&0&0\\0&0&0&0\\0&0&0&0
 \\0&0&0&-1\end {array} \right),\quad
 B_{E_3}=\left(\begin {array}{cccc} 0&0&0&0\\0&-1&0&0\\0&0&1&0
 \\0&0&0&0\end {array} \right),\quad
 B_{E_5}=\left(\begin {array}{cccc} 0&0&0&0\\0&0&0&0\\-1&0&0&0
 \\0&1&0&0\end {array} \right),\nn\\
 B_{E_8}=\left(\begin {array}{cccc} 0&0&0&0\\-1&0&0&0\\0&0&0&0
 \\0&0&1&0\end {array} \right),\quad
 B_{E_{10}}=\left(\begin {array}{cccc} 0&1&0&0\\0&0&0&0\\0&0&0&-1
 \\0&0&0&0\end {array} \right),\quad
 B_{E_{11}}=\left(\begin {array}{cccc} 0&0&1&0\\0&0&0&-1\\0&0&0&0
 \\0&0&0&0\end {array} \right).\nn
 \eea
The correspondence with the previously introduced generators is as
follows ($I=1,2,3,\;p=7,8$): \be P^I\leftrightarrow {E_I}, \quad
W_I\leftrightarrow { E_{I+3}}, \quad Z_I\leftrightarrow {E_{I+6}},
\quad \Omega^p\leftrightarrow {E_{p+3}},\quad X \leftrightarrow
{E_{12}}.\nn\ee In this representation, the matrices corresponding
to the negative roots, \be P^{-I}\leftrightarrow E_{-I}, \quad
W_{-I}\leftrightarrow E_{-(I+3)}, \quad Z_{-I}\leftrightarrow
E_{-(I+6)}, \quad \Omega^{-p}\leftrightarrow E_{-(p+3)},\quad X^{-}
\leftrightarrow E_{-12},\nn\ee are  transposed with respect to the
positive roots matrices:  \be
 E_{-k}=(E_k)^T.\nn
 \ee
The following normalization conditions are assumed: \be
 \texttt{tr}(H_i,H_j)=4\delta_{ij},\ i,j=1\ldots 4,\qquad
 \texttt{tr}(E_k,E_{-k})=2,\nn
 \ee
and the involution matrix $K$   is chosen as \be
 K=\rm{diag}(\kappa,\kappa,1,1,1,1,\kappa,\kappa).\nn
 \ee
The generators of the isotropy subgroup are selected by the Eq.
(\ref{h_K_h}). They are given by the following linear combinations
of the generators: \be
 P^I-\kappa P^{-I},\quad Z_I -\kappa Z_{-I},\quad W_I-W_{-I},\quad
 X-\kappa X^{-},\quad \Omega^7-\kappa \Omega^{-7},\quad
 \Omega^8-\Omega^{-8}.
\nn \ee

\section{Details of the coset matrices}

The block matrices entering $S_\Psi$ and $R_\Psi$, being expressed
though the target space potentials, read:
\bea \label{pssi}
 &&\e^{\frac12\Psi_{3}}\widetilde{\Psi}_a^T=
 \left( \begin{array}{ccc}
 \frac12(u^av^3-u^3v^a)\\
 -v^a\\
  -u^a\end{array} \right),\;\; a=1,2;\quad
 \Psi_{12}=
 -\left( \begin{array}{ccc}
 0&0&0\\
 v^1u^2&v^1v^2&0\\
 u^1u^2&u^1v^2&0  \end{array} \right),\nn\\
 &&\Psi_{3}\Psi_{12}
+\Psi_{12}\widetilde{\Psi}_{3}=
 \left( \begin{array}{ccc}
 v^3u^1u^2-u^3v^1u^2,&-u^3v^1v^2+v^3v^2u^1&0\\
 0&0&u^3v^1v^2-v^3v^2u^1\\
 0&0&-v^3u^1u^2+u^3v^1u^2  \end{array} \right),\\
 && \Psi_{3}\Psi_{12}\widetilde{\Psi}_{3}=
 \left( \begin{array}{ccc}
 0&0&(u^3)^{2}v^1v^2-u^3v^3v^1u^2-u^3v^3v^2u^1+(v^3)^2u^1u^2\\
 0&0&0\\
 0&0&0  \end{array} \right).\nn
  \eea
The explicit form for ${\cal Q}$ is:
\be
 {\cal Q}=\left( \begin{array}{cccc}
 \mu_1+ \frac{u_3v_2-v_3u_2}{2}&\omega_7-
 \frac{u_3v_1u_2-2u_3v_2u_1+v_3u_1u_2}{6}-u_2\mu_2,&
 \omega_8+\frac{v_3u_1v_2-2v_3u_2v_1+u_3v_1v_2}{6}-v_2\mu_2&0\\
  -v_2&-\mu_3+\frac{v_1u_2-u_1v_2}{2}&0&\\
   -u_2&0&&\\
    0&&&\\
 \end{array} \right),\nn
 \ee
 and the blocks entering $\cal P$ are
 \bea
  &&\e^{\Psi_3^T}\Lambda=
 \left( \begin {array}{cc}
 \kappa \e^{\sqrt2\phi_1}&0\\
 \kappa \e^{\sqrt2\phi_1}\eta&\widetilde\lambda^0
 \end {array} \right),\quad
 \eta=\left( \begin{array}{cc}
 u^3\\
  -v^3 \end{array} \right),\quad
 a=1,2\nn\\
 &&\e^{\Psi_3^T}\Lambda \e^{\Psi_3}=\kappa \e^{\sqrt2\phi_1}
 \left( \begin {array}{cc}
 1\\
 \eta
 \end {array} \right)\otimes\left(1,\eta^T\right)+
 \left( \begin {array}{cc}
 0&0\\
 0&\widetilde\lambda^0
 \end {array} \right).\nn
  \eea

\section{Transformations preserving asymptotic flatness}
In this appendix we exhibit the action of the transformations
generated by linear combinations of generators \be
 Z_I+Z_{-I},\quad W_{-I}+P^{-I},\quad X-\Omega^{-8}\nn
 \ee
on a neutral seed \vspace{-.3cm}:
\bea
 && X^1=X^2=X^3=1,\nn\\
 && \phi_1=\frac{1}{\sqrt2}(\varphi_2+\frac{1}{\sqrt3}\varphi_1),\quad
 \phi_2=\frac{1}{\sqrt2}(\varphi_2-\frac{1}{\sqrt3}\varphi_1),\quad
 \phi_3=\frac{\sqrt6}{3}\varphi_1,\quad  \phi_4=0,\nn\\
 && \Lambda=\left( \begin {array}{cc}
 -\tau^{-1}&0\\
 0&\widetilde\lambda\end {array} \right),\quad \lambda=\lambda^0\neq0,\quad
 \tau=\tau_0=\e^{-\sqrt2\phi_1},\quad  \omega_p\neq 0,\nn\\
 && \psi^I=0,\quad
 \mu_I=0.\nn
 \eea
The relation $e^{-\sqrt2\phi_1}=\tau$ is a consequence of the
condition $X^3=1$ (see Eq.(\ref{X_via_phi})). Such a  seed may
describe  a neutral black ring or a black hole with one or two
independent rotation parameters. Three above transformations
preserve asymptotic flatness with the Guisto-Saxena choice of
coordinates and generate some combinations of charges.  For this
seed the coset blocks simplify to
 \be
  {\cal P}_0=\left( \begin{array}{cc}
 \Lambda&0\\
 0&1
 \end {array} \right),\quad
 {\cal Q}_0=\left( \begin {array}{cc}
 0&\Omega\\
 0&0
 \end{array} \right),\nn
  \ee
and we obtain he following transformations of the potentials:
\vspace{-.2cm}
\subsection{$Z_I+Z_{-I}$} \vspace{-.7cm}
\be
 \lambda_{77}'= D^{-2/3}(\lambda_{77}c^2-\tau s^2),\quad
 \lambda_{78}'= D^{-2/3}c\lambda_{78},\quad
 \lambda_{88}'= D^{-2/3}\lambda_{88},\quad \tau'= \tau D^{-1/3},\nn
 \ee
\be
 \e^{\sqrt2\phi_1'}=\begin{cases}
 \tau^{-1},&\ I=1,2\\
 D\tau^{-1},&\ I=3\\ \end{cases},\quad
 \e^{\sqrt2\phi_4'}=\begin{cases}
 D^{-1},&\ I=1\\
 D &\ I=2\\
 1,&\ I=3\\ \end{cases}\nn
 \ee
\be
 (v^J)'=\delta^{JI}\frac{sc(1+\lambda_{88})}{D},\quad (u^J)'=
 \delta^{JI}\frac{s\lambda_{78}}{D},\quad
 (X^J)'=D^{-2/3}\delta^{IJ}+D^{1/3}(1-\delta^{IJ}).\nn
 \ee
\be
 \mu_I'=0,\quad \omega_7'=\omega_7,\quad \omega_8'=c\omega_8,\nn
 \ee
with
\be
 D=c^2+s^2\lambda_{88},\quad c\equiv \cosh(\alpha),\quad
 s\equiv \sinh(\alpha).\nn
 \ee \vspace{-.7cm}
\subsection{$W_{-I}+P^{-I}$} \vspace{-.7cm}
\be
 \widetilde{\lambda}_{77}'=\begin{cases}
 \widetilde{\lambda}_{77},&\ I=1,2\\
 \frac{D_2}{D_1}\widetilde{\lambda}_{77},&\ I=3\\ \end{cases},\quad
 \widetilde{\lambda}_{78}'=\begin{cases}
 \widetilde{\lambda}_{78}(1-\xi_2)-\widetilde{\lambda}_{77}\xi_3,&\ I=1,2\\
 \frac{D_2}{D_1}(\widetilde{\lambda}_{78}(1-\xi_2)-
 \widetilde{\lambda}_{77}\xi_3),&\ I=3\\ \end{cases},\nn
 \ee
 \be
 \widetilde{\lambda}_{88}'=\begin{cases}
 \widetilde{\lambda}_{77}\xi_3^2-2\widetilde{\lambda}_{78}
 \xi_3(1-\xi_2)+\widetilde{\lambda}_{88}(1-\xi_2)^2
 +\alpha^2(1-\frac12\xi_1),&\ I=1,2\\
 \frac{D_2}{D_1}(\widetilde{\lambda}_{77}\xi_3^2-
 2\widetilde{\lambda}_{78}\xi_3(1-\xi_2)+\widetilde{\lambda}_{88}(1-\xi_2)^2
 +\alpha^2(1-\frac12\xi_1)),&\ I=3\\ \end{cases},\nn
 \ee

\be
 \tau'=\begin{cases}
 \tau D_1^{-1},&\ I=1,2\\
 D_1 D_2^{-2}\tau,&\ I=3\\ \end{cases},\quad
 \e^{\sqrt2\phi_1'}=\begin{cases}
 D_2\tau^{-1},&\ I=1,2\\
 D_1\tau^{-1},&\ I=3\\ \end{cases},\quad
 \e^{\sqrt2\phi_4'}=\begin{cases}
 D_2/D_1,&\ I=1\\
 D_1/D_2,&\ I=2\\
 1,&\ I=3\\ \end{cases}\nn
 \ee

 \bea
 && \omega_7'=D_2^{-1}(\omega_7(1+\xi_2)-\lambda_{77}\xi_1),\quad
  \omega_8'=D_2^{-1}(\omega_8(1+\xi_2)-\lambda_{78}\xi_1),\nn\\
 && (v^J)'=\delta^{JI}D_1^{-1}\alpha(\lambda_{77}\xi_3+\lambda_{78}(1+\xi_2-\xi_1)-
 \omega_8(1-\xi_2)),\nn\\
 && (u^J)'=\delta^{JI}D_1^{-1}\alpha(\lambda_{77}(1-\xi_1)-\omega_7(1-\xi_2)),\nn\\
 && \mu_J'=\delta_{JI}D_2^{-1}\alpha(\omega_7(1+\xi_2)-\lambda_{77}\xi_1-\tau),\nn\\
 && (X^J)'=\Big(\frac{D_2}{D_1}\Big)^{-2/3}\delta^{IJ}+
 \Big(\frac{D_2}{D_1}\Big)^{1/3}(1-\delta^{IJ}).\nn
 \eea
where \vspace{-.5cm}

 \bea
 && \xi_1=\frac{\alpha^2\tau}{2},\quad \xi_2=\frac{\alpha^2\omega_7}{2},\quad
 \xi_3=\frac{\alpha^2\omega_8}{2},\quad \xi_4=\frac{\alpha^2\lambda_{77}}{2}.\nn\\
 && D_1=(1-\xi_2)^2+2\xi_4(1-\frac12\xi_1),\quad
 D_2=(1+\xi_2)^2-2\xi_1(1+\frac12\xi_4).\nn
 \eea \vspace{-.7cm}
\subsection{$X-\Omega^{-8}$} \vspace{-.7cm}
 \bea
 &&\widetilde{\lambda}_{77}'=-\alpha^2\tau+\widetilde{\lambda}_{77}\xi_1^2
 -2\widetilde{\lambda}_{78}\alpha\omega_7\xi_1+\widetilde{\lambda}_{88}
 \alpha^2\omega_7^2,\nn\\
 &&
 \widetilde{\lambda}_{78}'=\frac12\alpha^3\tau-\frac12\widetilde{\lambda}_{77}
 \alpha\xi_1(1+\xi_1)
 +\widetilde{\lambda}_{78}(\xi_1+\xi_2+2\xi_1\xi_2)
 -\widetilde{\lambda}_{88}\alpha\omega_7(1+\xi_2),\nn\\
 &&\widetilde{\lambda}_{88}'=-\frac14\alpha^4\tau+\frac14\widetilde{\lambda}_{77}
 \alpha^2(1+\xi_1)^2
 -\widetilde{\lambda}_{78}\alpha(1+\xi_1)(1+\xi_2)+\widetilde{\lambda}_{88}(1+\xi_2)^2,\nn\\
  && \phi_4'=0,\quad \e^{\sqrt2\phi_1'}=D_{\omega}e^{\sqrt2\phi_1},\quad
 \tau'=D_{\omega}^{-1}\tau,\nn\\
  &&\omega_7'=D_{\omega}^{-1}\left(\omega_7(\xi_1-\xi_2)
 +\alpha(\frac12\alpha\lambda_{77}+\lambda_{78})\tau\right),\nn\\
 &&\omega_8'=D_{\omega}^{-1}\left((\omega_8+\alpha\omega_7)(\xi_1-\xi_2)+
 \alpha(\frac12\alpha^2\lambda_{77}
 +\frac32\alpha\lambda_{78}+\lambda_{88})\tau\right),\nn
 \eea
  \bea
 && (\psi^I)'=0,\quad \mu_I'=0,\quad (X^I)'=1,\nn\\
 &&D_{\omega}=(1-\alpha\omega_8-\frac12\alpha^2\omega_7)^2
 -\alpha^2(\frac14\alpha^2\lambda_{77}+\alpha\lambda_{78}+\lambda_{88})\tau,\nn\\
 &&\xi_1=1-\alpha\omega_8,\quad \xi_2=\frac12\alpha^2\omega_7.\nn
   \eea

\section{Potentials in terms of ${\cal P}$ and ${\cal Q}$}
Here we give the explicit expressions for the    target space
potentials in terms of the components of the matrix ${\cal P}$ and
${\cal Q}$:
\be
 \psi^1=\frac{1}{{\cal P}_{11}D_{11,22,33}}
 \left( \begin {array}{cc}
 D_{11,23}D_{11,24}-D_{11,22}D_{11,34}\\
 D_{11,23}D_{11,34}-D_{11,24}D_{11,33}
  \end {array} \right),\quad
  \psi^2=
 \left( \begin {array}{cc}
 {\cal Q}_{42}\\
 {\cal Q}_{43}
  \end {array} \right),\quad
 \psi^3=
 \left( \begin {array}{cc}
 {\cal P}_{12}/{\cal P}_{11}\\
 -{\cal P}_{13}/{\cal P}_{11}
  \end {array} \right),\nn
 \ee
where $D_{ij,kl}$ denotes the determinant of $2\times 2$ matrix
constructed from ${\cal P}_{ij}$:
\be
 D_{ij,kl}={\cal P}_{ij}{\cal P}_{kl}-{\cal P}_{ik}{\cal P}_{jl},\nn
 \ee
and $D_{11,22,33}$ is the determinant of the $3\times 3$ minor of
the $4\times 4$ matrix ${\cal P}$ with the diagonal ${\cal
P}_{11},{\cal P}_{22},{\cal P}_{33}$. The remaining quantities read:
\be
 \kappa \e^{\sqrt2 \phi_1}={\cal P}_{11},\quad \widetilde\lambda^0=
 \frac{1}{{\cal P}_{11}}\left( \begin {array}{cc}
 D_{11,22}&D_{11,23}\\
 D_{11,23}&D_{11,33}
 \end {array} \right),\nn
 \ee
\bea
 \mu_1&=&{\cal Q}_{11}-\frac{1}{2{\cal P}_{11}}({\cal P}_{12}{\cal Q}_{43}+
 {\cal P}_{13}{\cal Q}_{42}),\nn\\
 \mu_2&=&\frac{{\cal P}_{14}}{{\cal P}_{11}}+\frac12(u^3v^1-v^3u^1),\nn\\
 \mu_3&=&{\cal Q}_{33}+\frac12(v^1u^2-v^2u^1).\nn
 \eea
 \be
 \e^{\sqrt2\phi_4}={\cal P}_{44}- (\mu_2+\frac12(v_3u_1-v_1u_3))^2{\cal P}_{11}-
 \widetilde\lambda_{77}^0v_1^2- 2
 \widetilde\lambda_{78}^0v_1u_1-\widetilde\lambda_{88}^0u_1^2,\nn
   \ee
   \bea
   \omega_7&=&{\cal Q}_{12}+u_2\mu_2+\frac16 u_1u_2v_3-\frac13 u_3u_1v_2+\frac16
   u_3u_2v_1,\nn\\
  \omega_8&=&{\cal Q}_{13}+v_2\mu_2-\frac16 u_3v_1v_2+\frac13 v_3v_1u_2-\frac16
   v_3v_2u_1.\nn
    \eea


\end{document}